\documentclass[9pt,sigconf]{acmart}

\copyrightyear{2026}
\acmYear{2026}
\setcopyright{cc}
\setcctype{by}
\acmConference[IMC '26]{Proceedings of the 2026 ACM Internet Measurement Conference}{October 12--16, 2026}{Karlsruhe, Germany}
\acmBooktitle{Proceedings of the 2026 ACM Internet Measurement Conference (IMC '26), October 12--16, 2026, Karlsruhe, Germany}
\acmDOI{10.1145/3777912.3809139}
\acmISBN{979-8-4007-2327-8/2026/10}

\usepackage{subcaption}[tight]
\usepackage{fancyvrb}
\usepackage{xspace}
\usepackage{enumitem}
\usepackage{cleveref}
\usepackage{pifont}

\newcommand{\xmark}{\ding{55}}%
\usepackage{tablefootnote}
\usepackage{multirow}

\newcommand{\ie}{i.e., \@}
\newcommand{\eg}{e.g., \@}

\newcommand{\etal}{et al.\xspace}

\newcommand{\punkt}[1]{\item\textbf{#1}:}
\newcommand{\afrinic}{AFRINIC\xspace}

\graphicspath{{figures/}}

\usepackage[absolute,showboxes]{textpos}

\begin{document}

\settopmatter{printfolios=true}
\setlength{\TPHorizModule}{\paperwidth}
\setlength{\TPVertModule}{\paperheight}
\TPMargin{5pt}
\begin{textblock}{0.8}(0.1,0.02)
     \noindent
     \footnotesize
     If you cite this paper, please use the ACM IMC 2026 reference:
     Robert Beverly, Amreesh Phokeer, and Oliver Gasser. 2026.
     WHEREIS: IP Address Registration Geo-Consistency.
     In \textit{Proceedings of the 2026 ACM Internet Measurement Conference (IMC ’26), October 12–16, 2026, Karlsruhe, Germany.}
     ACM, New York, NY, USA, 15 pages. https://doi.org/10.1145/3777912.3809139
\end{textblock}
 
\title{WHEREIS: IP Address Registration Geo-Consistency}

\author{Robert Beverly}
\affiliation{%
  \institution{San Diego State University}
  \country{}
}
\email{rbeverly@sdsu.edu}

\author{Amreesh Phokeer}
\affiliation{%
  \institution{Internet Society}
  \country{}
}
\email{phokeer@isoc.org}

\author{Oliver Gasser}
\affiliation{%
  \institution{IPinfo}
  \country{}
}
\email{oliver@ipinfo.io}

\ccsdesc[500]{Networks~Network measurement}
\ccsdesc[500]{Networks~Network management}

\keywords{WHOIS, RIR, IP Registration, IP Geolocation}

\begin{abstract}
The five Regional Internet Registries (RIRs) provide the critical
function of IP address resource delegation and registration.  The
accuracy of registration data directly impacts Internet operation, 
management, security, and optimization.  In addition, the scarcity
of IP addresses has brought into focus conflicts between 
RIR policy and IP registration ownership and use.  The tension
between a free-market based approach to IP address allocation 
and use versus
policies to promote fairness and regional equity has resulted in
litigation and debate that threatens the existence of the RIR system.

We develop WHEREIS, a measurement-based system to geolocate
delegated IPv4 and IPv6 prefixes at an RIR-region granularity, and
systematically study \emph{where} addresses are physically used
\emph{post-allocation} as well as the extent to which registration
information is accurate.  We define a taxonomy of registration
``geo-consistency'' that compares a prefix's measured geolocation 
against the allocating RIR's coverage region and the registered
organization's location.  While in aggregate over 98\% of the prefixes
we examine are consistent with our geolocation inferences, there is
substantial variation across RIRs.  
We find that
IPv6 registrations are no more consistent than
IPv4, suggesting that structural, rather than technical, issues play
an important role in allocations.  
We solicit additional information
on inconsistent prefixes from network operators, IP leasing providers,
and collaborate with three RIRs to obtain validation.  
We further show
that the inconsistencies we discover manifest in three commercial
geolocation databases.  Finally, we focus on \afrinic as a case study 
of particular importance. 
By improving the transparency around
post-allocation prefix use, we hope to improve applications that use
IP registration data and inform ongoing discussions over in-region
IP address use and policy.
\end{abstract}

\maketitle

\section{Introduction}
\label{sec:intro}

To ensure the global uniqueness of public IP
addresses, the IANA allocates
large contiguous blocks of addresses to 
Regional Internet
Registries (RIRs) who then further assign or re-allocate within their respective
geographic region.  

There are currently five RIRs, each of which serves a different
geographical region of the world.  
Assignments, allocations, and reassignments are governed by RIR
policies \cite{arin_policy,lacnic_policy,apnic_policy,ripe_policy,afrinic_policy} that are designed to provide uniqueness, efficiency, and
accountability for their respective regions.  
With the advent of IPv4 address
exhaustion \cite{richter2015primer}, these policies have become
increasingly important as address space is now a valuable commodity
\cite{prehn2020wells}.
The intent of regional delegation 
is explicitly codified by the ICP-2 document on the establishment of a new RIR: 
``The proposed RIR must operate internationally in a large geographical region of approximately continental size'' \cite{icp2}.

While the canonical view is that networks located within a particular
region obtain IP addresses from the RIR responsible for that region,
operational practice is often much more
complex.  For instance, a company in the Americas might obtain addresses
from the European RIR, but use those addresses for a datacenter in
Asia.  Such ``inconsistencies'' may be attributable to business,
political, abusive, or legal factors, or may simply be due to operational
expedience.  The tension between a free-market based approach to 
address allocation versus more needs-based policies to promote 
fairness and regional equity has become especially contentious with
recent litigation threatening the very existence of the
RIR system~\cite{afrinic-faq, afrinic-court}.  From an operational
perspective, prefixes that are geographically inconsistent with their 
registration
data can impede network monitoring, security, and management.

In
this work, we take a rigorous, measurement-based look at the
current state of address registration ``geo-consistency'' across the
five RIRs to:
i) quantify the extent to which registry information
is accurate and can serve operational needs; 
ii) increase transparency
and help the community better understand where addresses are 
being used; 
and iii) inform
ongoing discussion and current debate over out-of-region (OOR) address
use and policy~\cite{afrinic-faq, afrinic-court, ci-seized,
sa-heist, krebs}.
The primary contributions of this work include:

\begin{itemize}[leftmargin=*]
    \item \textbf{Analysis of IP registration in WHOIS
     (\S\ref{sec:data}):}
       Across the five RIRs we find 
       variation (0.2--15\%) in OOR
       organization registrations between RIRs, and 
       fewer OOR IPv6 prefixes as 
       compared to IPv4.
   \item \textbf{Geo-consistency taxonomy
       (\S\ref{sec:method}):}
       We create a taxonomy of prefix registration geo-consistency, covering different characteristics such as RIR location, registered organization location, and physical location.
       We leverage this taxonomy to identify different levels of geo-inconsistencies using Internet measurements.
   \item \textbf{IPv4 and IPv6 prefix geo-audit
       (\S\ref{sec:results}):}
     We perform an active measurement %
       campaign and devise a methodology to determine
       whether prefixes are used in a physical location 
       consistent with the RIR's region and registered 
       organization's country.  While more than 98\% of IPv4 and 
       IPv6 prefixes are fully consistent, we identify hundreds
       of inconsistent prefixes and find substantial variation
       between RIRs, with AFRINIC exhibiting the highest fraction
       of inconsistencies in IPv4 and APNIC and LACNIC in IPv6.
   \item \textbf{Analysis of properties for consistent and
inconsistent prefixes
       (\S\ref{sec:results:properties}):}
       To understand possible contributing factors to
       inconsistent prefixes, we examine prefix age, size,
       and type, and discover that most inconsistencies in 
       IPv4 are attributable to legacy prefixes, while 
       inconsistent IPv6 prefixes are more likely to be
       recently updated.
   \item \textbf{Reproducibility and impact
       (\S\ref{sec:impact})}:
       We make our code and measurement data publicly available
       to facilitate reproducibility.  Further,
       we shared our inconsistent prefix inferences with three
       RIRs who are using our findings to inform their registration
       database and process. 
\end{itemize}

\noindent By improving the
transparency around this aspect of post-allocation prefix use, our
hope is to improve applications that use IP registration data and
inform ongoing discussions over in-region address use and policy---a
crucial aspect of a core component of the Internet's future.
\section{Background}
\label{sec:background}

IP addresses are fundamental to the Internet's
operation.  The Internet numbers registry system has three primary
goals for IP addresses: (1) allocation pool management, (2) hierarchical
allocation, and (3) registration accuracy~\cite{rfc7020}.  The IP
address hierarchy is rooted in the Internet Assigned Numbers Authority
(IANA) which is managed by the Internet Corporation for Assigned Names
and Numbers (ICANN) organization.  RFC 1366 first proposed to
geographically distribute the registry functionality~\cite{rfc1366}.
Today IANA allocates large (\eg /8's in IPv4 and /12's in IPv6), 
unique blocks of IP address space to the
Regional Internet Registries (RIRs), of which there are currently
five: ARIN (North America),
RIPE (Europe, the Middle East, and parts of Central Asia), APNIC (Asia and Pacific), LACNIC (Latin America and the Caribbean), or
\afrinic (Africa and the Indian Ocean).  RIRs then further allocate IP address blocks to LIRs (Local Internet Registries such as an Internet Service Provider (ISP) or to NIRs (National Internet Registries), which are organizations responsible for the management of IP address allocations at national level (e.g., CNNIC in China).
This distribution affords autonomy to the
different RIRs to consider region-specific geopolitical policies and
constraints, as well as attempting to balance concerns of fairness
and regional resource equity.

\subsection{Motivation}

\vspace{2mm}\noindent\textbf{Policy:}
Our work is motivated by the ongoing crisis within the AFRINIC region 
where significant debate and even legal actions---which may threaten
the ongoing viability of the RIR itself---following AFRINIC's attempts
to enforce in-region address  use. AFRINIC has faced prolonged legal challenges  since 2020, primarily triggered by disputes over IP address allocations and  allegations of governance failures. The most notable conflict involves Cloud Innovation Ltd., which sued AFRINIC after the registry attempted to reclaim large blocks of IP addresses were allegedly not being utilized for the purposes for which they were issued. The legal battles have led to bank account freezes, operational disruptions,  and governance paralysis, raising concerns about Internet number resource stability and RIR accountability  in the region.~\cite{afrinic-court,afrinic-faq,ci-seized,sa-heist,krebs,crumbs,arin_afrinic_recap}.

As noted in the ARIN Number Resource Policy Manual: ``The primary role of RIRs is
to manage and distribute public Internet address space within their
respective regions'' \cite{arin_policy}.  However, RIRs must balance
efficient and equitable use of IP addresses with the true need for
addresses, as well as real-world operational constraints.  Further, 
the five RIRs have \emph{different} policies with respect to 
out-of-region address use; Table~\ref{tab:nro} summarizes pertinent
policies from NRO's Comparative Policy Overview~\cite{nro}. 

AFRINIC has the most restrictive out-of-region usage policies (currently being in the IPv4 exhaustion phase). The account holder must be legally present in the service region and the infrastructure from which the services are originating must be located in
the AFRINIC service region. Out-of-region usage is only allowed if the purpose is to bring back connectivity to Africa.
ARIN's policy is quite restrictive. Applicants using IP addresses out-of-region should demonstrate a real and
substantial connection with the ARIN region \cite{arin_policy}.
On the other hand, APNIC and RIPE have the most relaxed out-of-region usage policies and they explicitly allows out-of-region use without restriction for its members.
\begin{table}[]
       \caption{NRO Comparative Policy Overview: membership and out-of-region (OOR) policies across RIRs \cite{nro}.
                 A \checkmark\xspace indicates ``allowed'', while an
                 \xmark\xspace indicates ``against'' policy.  OOR
                 polices vary significantly across RIRs.}
	\label{tab:nro}
    \resizebox{\columnwidth}{!}{%
	\begin{tabular}{c|ccc|c}
		\toprule
		\multirow{2}{*}{} & \multicolumn{3}{c|}{Out-of-region}                                & \multirow{2}{*}{\begin{tabular}[c]{@{}c@{}}Out-of-region\\ usage\end{tabular}} \\ \cline{2-4}
		& \multicolumn{1}{c|}{Owner} & \multicolumn{1}{c|}{Infrastructure} & Routing &                                                                                \\ \hline
		AFRINIC           & \multicolumn{1}{c|}{\xmark}     & \multicolumn{1}{c|}{\xmark}              & \xmark       & \xmark\tablefootnote{Only allowed for connectivity back to Africa.}                                                                    \\
		APNIC             & \multicolumn{1}{c|}{\checkmark}     & \multicolumn{1}{c|}{\checkmark}              & \checkmark       & \checkmark                                                                              \\
		ARIN              & \multicolumn{1}{c|}{\xmark}     & \multicolumn{1}{c|}{\xmark}              & \xmark       & \xmark\tablefootnote{Must prove real and substantial connection with the ARIN region.}                                                                    \\
		LACNIC            & \multicolumn{1}{c|}{\xmark}     & \multicolumn{1}{c|}{\xmark}              & \checkmark       & \xmark\tablefootnote{Allowed for Anycast prefixes}                                                                    \\
		RIPE              & \multicolumn{1}{c|}{\checkmark}     & \multicolumn{1}{c|}{\xmark}              & \checkmark       & \checkmark                                                                              \\ 
			\bottomrule
	\end{tabular}
}
\end{table}

\vspace{2mm}\noindent\textbf{Network Operations:}
Our work has been performed in coordination with 
three RIRs: ARIN, RIPE, and AFRINIC.  As discussed
in~\S\ref{sec:impact}, we share our data with representatives
at these RIRs and obtain informal validation.  Our work is further
informed by input
from the network operator community at NANOG and RIPE meetings. 
We have received consistent feedback from the
operational community that our work is vital to their applications,
work, and policies.  Indeed, IP registration data is widely used 
in network monitoring, security, and management.  As one individual
at an RIR commented to us: ``As the country-codes by RIRs are by far
the easiest way to group resources by country, your analysis sheds
light on to what extent this is accurate.''  

\vspace{2mm}\noindent\textbf{Database Maintenance:}
Network operators rely on the WHOIS as a database of record for
coordination, business continuity, and security operations.
An employee at one
RIR confirmed that, while they perform pre-delegation checks for
policy and region compliance, they have no mechanisms or regular
procedures to check for compliance post-allocation---only performing
manual investigations upon receiving a complaint.  
In addition, countless applications, security appliances, and
location services rely on WHOIS registration data to provide coarse-grained
geolocation information; Section~\ref{sec:impact:geodb} examines the
impact of geo-inconsistent prefixes on three commercial geolocation
databases.

\subsection{Terminology}

Each RIR maintains registration information, including for example the
assigned organization, mailing address, and points of contact, for
numbered resources (Figure~\ref{fig:record} provides an example).  
Accurate registration information is important to
the operation and management of the 
Internet---for instance to facilitate timely coordination during a
network attack or outage~\cite{nobile}.  This registry
information is exposed via a public directory service known as ``WHOIS''
~\cite{rfc3912}.

There are two primary types of IP address registration. The first is an allocation, for
blocks of IP addresses which will later be reallocated or reassigned
to third parties. The second is an assignment, for blocks that will
not be reassigned, or the recipient is an end-user or end-site.
Registries such as ARIN require
the reassignment 
of IPv4 prefixes of /29 or more addresses to be 
registered via a directory services system such as Shared WHOIS
(SWIP) within seven days~\cite{rfc1491,ripeswip}.  

\noindent In this work, we define two regional
locations:
\begin{itemize}[leftmargin=*]
 \punkt{Out of Region Organization (OOR)} If the registered prefix owner
        organization's physical contact address (mailing address)
        is in a country that is not part of the RIR's region, we
        say this is an Out of Region Organization (OOR). 
 \punkt{Out of Region Prefix (ORP)} When our 
        geolocation of a prefix places the IP addresses outside 
        the registered RIR's region, we call this an Out of Region
        Prefix (ORP).
\end{itemize}

As the registered owner of an address prefix may be 
a multi-national company or the customer of an ISP, 
we take a conservative view of what constitutes an inconsistency by 
considering both whether a prefix is OOR and/or ORP.
This allows us to define different categories
of inconsistency (see taxonomy in~\Cref{sec:method:taxonomy}) and focus on 
instances where the registrations are 
most inconsistent.

\begin{figure}[!bpt]
\centering
{\scriptsize
\begin{Verbatim}[frame=single]
NetHandle:      NET-104-148-63-0-1
OrgID:          C05266659
Parent:         NET-104-148-0-0-1
NetName:        WEB-OMEGA-DO-BRASIL
NetRange:       104.148.63.0 - 104.148.63.255

OrgID:          C05266659
OrgName:        Web Omega do Brasil
Street:         Rua do Xareu, qd 13, lote 20
City:           Goiania
State/Prov:     GO
Country:        BR
\end{Verbatim}
}
\vspace{-4mm}
\caption{Example ARIN prefix registration record.
The corresponding organization's country, Brazil,
is within the region of a different RIR.  WHEREIS
seeks to resolve this by geolocating the prefix's true location.
}
\vspace{-4mm}
\label{fig:record}
\end{figure}

\begin{table*}[!ht]
\caption{WHOIS dataset (Oct 1, 2024), including
proportion of prefixes and addresses with Out of Region Owners.}
\label{tab:overview}
\vspace{-2mm}
{\small
\resizebox{2.0\columnwidth}{!}{
\begin{tabular}{l|rr|rr|rr|rr}
\toprule
    & \multicolumn{4}{|c|}{IPv4} &\multicolumn{4}{c}{IPv6} \\
\cmidrule(lr){2-5}
\cmidrule(lr){6-9}
RIR &                 & Out-region         & Addresses & Out-Region 
    &                 & Out-region         & Addresses & Out-Region \\
    & Prefix (k)      & Owner (k prefix) & (M /24s)    & Owner (M /24s)
    & Prefix (k)      & Owner (k prefix) & (M /48s)    & Owner (M /48s) \\
\midrule
ARIN & 3,104.4 & 70.8 (2.3\%) & 5.2 & 0.1 (2.5\%)
     & 308.6 & 2.3 (0.7\%) & 13342.2 & 1.8 (0.0\%)\\

RIPE & 3,275.9 & 23.4 (0.7\%) & 2.8 & 0.0 (1.5\%)
     & 359.4 & 3.6 (1.0\%) & 6993.3 & 80.8 (1.1\%)\\

APNIC & 1,187.3 & 3.2 (0.3\%) & 9.4 & 0.0 (0.1\%)
      & 108.2 & 1.2 (1.1\%) & 7443.5 & 25.4 (0.3\%)\\

LACNIC & 475.5 & 1.1 (0.2\%) & 0.7 & 0.0 (0.0\%)
       & 28.4 & 0.1 (0.4\%) & 1082.2 & 1.4 (0.1\%)\\

AFRINIC & 162.8 & 23.9 (14.7\%) & 0.5 & 0.0 (5.7\%)
        & 32.3 & 0.0 (0.1\%) & 715.6 & 0.0 (0.0\%)\\
\midrule
Total: & 8,205.9 & 122.4 & 18.6 & 0.2
       & 836.9 & 7.3 & 29576.9 & 109.4\\
\bottomrule
\end{tabular}}}
\end{table*}

\section{WHOIS Dataset}
\label{sec:data}

Before describing our geo-consistency measurement methodology 
(Section~\ref{sec:method}),
we first detail and examine important
characteristics of the prefix registration data in WHOIS.

We use raw ``bulk WHOIS'' dumps
of the IPv4 and IPv6 prefix registration databases as of 
October 1, 2024 from
all five RIRs.
Doing so 
ensures that we
obtain the complete data en masse~\cite{bulkwhois}.  Note that the
RIRs also publish publicly available delegation files that 
summarize in a standard format core information about the prefixes
they manage including status (allocated or assigned), the country code of the 
organization to which the prefix is delegated, and the date of
delegation~\cite{delegation}.  As described by Arouna \etal, delegation files can
be combined with publicly available WHOIS data, for instance to 
obtain the registration ``status'' value missing in public ARIN WHOIS
or the ``maintainer'' in public LACNIC WHOIS~\cite{arouna2023lowering}.  
We analyze the differences between the bulk WHOIS data from ARIN and
the corresponding public extended delegation statistics file from the
same date and find \eg less than 0.1\% disagreement in registered 
country code.
Thus, as we apply
for, and receive, research access to the complete (non-public) 
bulk WHOIS data from the
RIRs, we henceforth use the bulk WHOIS data in our analysis. 
Note that we focus on the five RIRs and exclude WHOIS data of
National Internet Registries (NIRs). NIRs usually have their own IP addressing and management policy which is outside the scope of this study. 

\subsection{Macro Characteristics}

The bulk WHOIS data is organized in flat text files
containing records that consist of key-value pairs, \eg ``NetRange''
and ``Country'' as shown in Figure~\ref{fig:record}.  Each RIR's database has different
schemas and idiosyncrasies, these include for example the address prefix 
representation, different key names, transferred prefixes, and 
prefixes from other RIRs.  We parse each prefix allocation and skip
prefixes the RIR does not manage (but still has listed in their
database for completeness \eg annotated with a ``not-managed-by'' note). 
We find a number of fine-grained anomalies in WHOIS records, which we deal with diligently (see \Cref{app:anomalous} for details).

Some RIRs do not include a country attribute in registration records,
but rather use an ``OrgID'' and include additional records for
organizations (see Figure~\ref{fig:record}).
When an OrgID exists, we map the organization's country to the prefix.
ARIN, RIPE, and AFRINIC include organization records, while APNIC and
LACNIC do not.  In total, we map countries for 3,130,262 prefixes
from their respective organization records; almost all of these are
within ARIN.

\begin{figure*}[t!]
 \centering
\begin{subfigure}{0.5\textwidth}
  \centering
  \includegraphics[width=.9\linewidth]{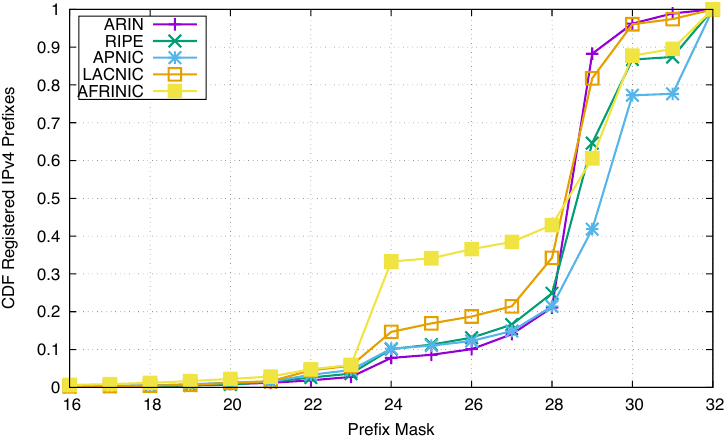}
\end{subfigure}%
\begin{subfigure}{.5\textwidth}
  \centering
  \includegraphics[width=.9\linewidth]{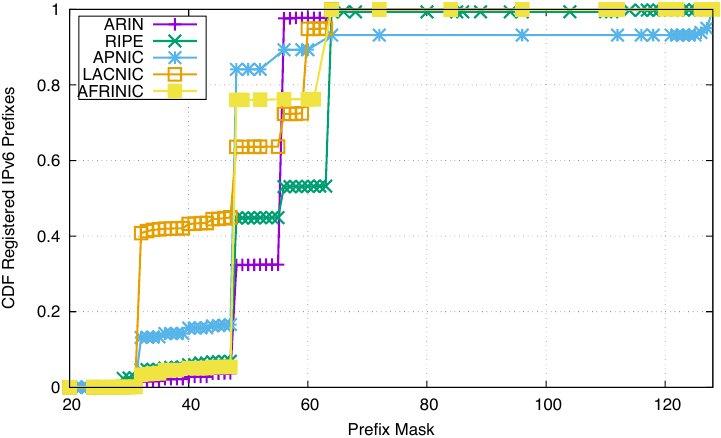}
\end{subfigure}
 \caption{Prefix length distribution of registered IPv4 and IPv6 prefixes per-RIR.}
 \label{fig:plen}
\end{figure*}

Our WHOIS data contains over 8.2M IPv4 and 0.8M IPv6 prefix registrations across the five RIRs,
representing a more granular allocation than in
the global BGP table \cite{hsu2023fiat}.  
Figure~\ref{fig:plen} depicts the cumulative
fraction of prefixes as a function of the prefix mask across the RIRs.
We find nearly 90\% of the registered IPv4 prefixes within ARIN, RIPE,
and APNIC have a prefix length of /25 or more specific, suggesting active
use of SWIP.  In contrast, a larger proportion of registered prefixes 
within \afrinic and 
LACNIC are less specific (\ie smaller network mask).
In IPv6, a majority of registrations across all RIRs have a prefix length
of /48 or more specific but the distributions of the prefix lengths
differ greatly among the RIRs. In particular, LACNIC has a substantial
prevalence of less specific registrations, with 45\% of prefixes being
less specific than /48, while in the rest of RIRs such prefixes are much
less prevalent, never reaching 17\% of registrations and, in the case
of ARIN, accounting for only 3.7\% of registrations.

\subsection{Inter-region Registration}
\label{sec:interregion}

Before performing active prefix geolocation in
Section~\ref{sec:method}, we first examine the location information
available within the bulk WHOIS data and analyze the prevalence of
Out of Region Organizations (OORs).

Prefixes are regularly transferred between RIRs; these transfers 
are logged and prior work has characterized such published
transfers~\cite{2017-livadariu-itmw}.  In contrast, this subsection
examines allocations in RIRs that are registered to 
organizations outside of the RIR's region.  For example, a
prefix $P$ belongs to an address block allocated to ARIN, but the
registered organization within the ARIN database has a physical
address in Brazil. In this case, the organization's country
belongs to the LACNIC region---a different RIR.  Figure~\ref{fig:record} provides
one such example WHOIS record.

\Cref{tab:overview} provides a macro-level analysis of 
the number of IPv4 and IPv6 prefixes (in thousands) in each RIR, as well as the fraction
of prefixes and addresses (in /24 equivalents) registered to
out-of-region (OOR) organizations.  Note that 
the sum of all addresses across the RIRs is larger than the total
IPv4 address space---this is due to overlap between prefixes within
different RIRs due to ``legacy'' registrations.

More
specifically, the LACNIC and AFRINIC RIRs are newer than the
other RIRs.  Historically, ARIN managed those resources in 
Latin America, while RIPE managed IP allocations assigned in
the African continent.  The so-called Early Registration 
Transfer (ERX) project was the concerted effort to transfer
these early registrations to these registries dedicated to 
new geographic regions~\cite{arinerx,ripeafrinic,erxlacnic}.

For example, \afrinic is responsible for 154.0.0.0/8,
however ARIN has a registered allocation of 154.1.0.0/16 (for the
company Goldman Sachs). This is a special case called Early
Registration Transfers (ERX), where initially the 154/8 was managed by
ARIN but got later transferred to \afrinic after the latter was
setup~\cite{arinerx}.

ARIN has an appreciable fraction (2.3\%) of prefixes with
out-of-region registered organizations, and dominates in 
total volume of both out-of-region registered prefixes and addresses.
Given the long history of ARIN and the ERX efforts to transfer
addresses, this finding is not wholly unsurprising.  

\afrinic is the most recently formed registry (2004). 
Nearly 15\% of \afrinic prefixes are allocated to
out-of-region organizations, suggesting significant 
apparent involvement of organizations outside of Africa and
possibly use of these resources 
outside of Africa (a hypothesis we
test in the next section).  Similarly, \afrinic leads in out-of-region
registrations when accounting by total IP addresses (5.7\%).

In IPv6 we also find stark differences in OOR between RIRs and
differences compared to IPv4.
We find APNIC to have the highest share of OOR registrations (1.1\%).
\afrinic, with 15\% of all IPv4 prefixes OOR, accounts for a meager 0.1\% in IPv6.
This hints at the use of African IP resources outside the continent for IPv4 only, possibly linked to the IPv4 address shortage \cite{richter2015primer,prehn2020wells}.
When looking at the share of OOR addresses normalized by /48 prefixes, we see that RIPE has a higher share of OOR addresses (1.1\%) compared to OOR prefixes (1.0\%), which indicates that this RIR's OOR prefixes are on average less-specific than /48 and thus contain a larger number of addresses.
Contrary, all other four RIRs have a lower share of OOR addresses compared to prefixes, showing the use of OOR prefixes more-specific than /48.

\subsection{BGP}
\label{sec:data:bgp}

\begin{figure}[] 
	\centering
	\resizebox{1.0\columnwidth}{!}{\includegraphics{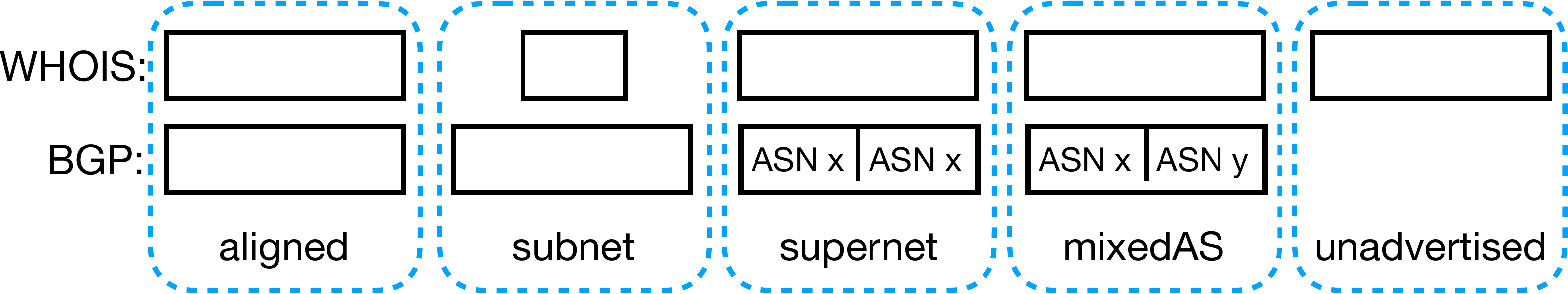}}
	\vspace{-1mm}
	\caption{A WHOIS prefix record can align with BGP or be a subnet,
		supernet, or a supernet containing BGP announcements from different
		ASNs.  In some cases, the WHOIS prefix may not be advertised in the
		BGP.}
	\label{fig:bgp}
	\vspace{-2mm}
\end{figure}

Our data analysis confirms previous findings that there are more
prefixes within the WHOIS as
compared to the BGP~\cite{hsu2023fiat}.
While this implies that many of the WHOIS
prefixes are subnets within a larger BGP aggregate, this is not always
the case.  In this subsection, we analyze the degree of alignment
(\ie the amount of address intersection; see Figure~\ref{fig:bgp})
between the WHOIS prefixes across the different RIRs with their
corresponding BGP announcement(s).

To perform this analysis, we lookup the WHOIS prefixes within a
snapshot of the global BGP table as captured on the same date as
the bulk WHOIS data by the Routeviews 
project~\cite{rv}.  In addition to a simple
longest-prefix match, we also enumerate the leaves of the best
matching node in the prefix trie for those prefixes that are contained
within the WHOIS prefix.  To see why this enumeration is necessary,
consider one WHOIS prefix for ``Sony Interactive'':
\texttt{100.42.96.0/20}.  While this prefix is not announced within
the BGP, there are 10 subnets of this prefix (of size \texttt{/23} and
\texttt{/24}).  The WHOIS prefix in this case is a supernet of the BGP
announcements, likely due to Sony's CDN making geographic-specific
announcements.  Our measurements support this inference, as a target
within one /23 is clearly within Europe, while a target within a 
different /23 of the /20 is clearly within the US.  Thus, it is
important to differentiate these instances of mis-alignment with
the BGP when analyzing our geo-audit results.

Table~\ref{tab:bgp} in \Cref{app:bgp} details the WHOIS to BGP alignment 
results---the most common case in IPv4, accounting for 
95\% of WHOIS prefixes, is when the WHOIS
prefix is a subnet of a larger BGP announcement. The next most
common case in IPv4 are WHOIS prefixes that are the same as the BGP prefix;
we term these ``aligned.''  \afrinic as the largest proportion of 
aligned prefixes, with nearly 9\% of their WHOIS entries exactly 
matching the BGP announcement.
The next case is supernetting, where the WHOIS prefix contains 
multiple BGP announcements.  While only 0.9\% of all WHOIS entries
are supernets from the same AS, LACNIC and \afrinic have a higher proportion with 
2.9\% and 2.4\% respectively. 
Finally, it can be the case that 
the BGP announcements within the WHOIS supernet are originated by
different autonomous systems (ASes)---we term this case ``MixedAS''---which occurs for 0.3\% of all WHOIS prefixes.

To eliminate cases where inconsistent geolocation for a WHOIS prefix
is due to BGP subnetting, we restrict the remainder of our analysis to 
WHOIS prefixes that are aligned with, or a subnet of, their BGP
counterpart. 
\begin{figure}[t!] 
	\centering
	\resizebox{0.6\columnwidth}{!}{\includegraphics{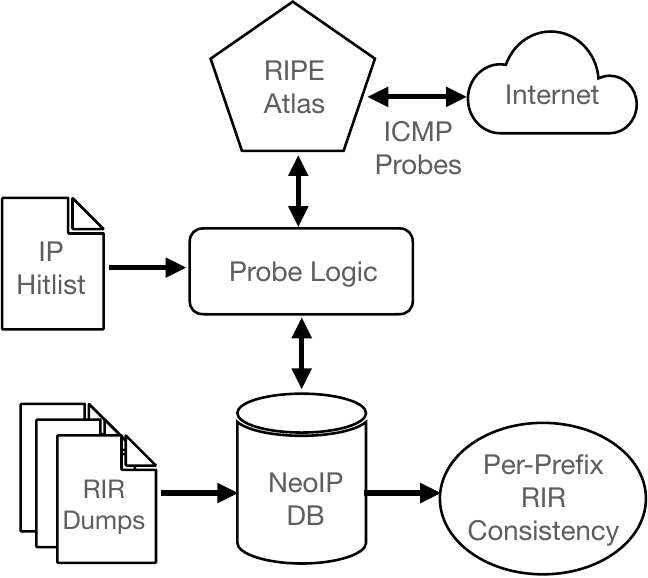}}
	\caption{WHEREIS methodology:  RIR prefixes are ingested
		into a database. Probing logic uses IPv4 and IPv6 hitlists 
                to drive RIPE Atlas active probing for constraint-based
                geolocation and RIR consistency checks.}
	\label{fig:method}
	\vspace{-4mm}
\end{figure}

\section{WHEREIS Prefix Geo-Audit Methodology}
\label{sec:method}

Having examined prefix registrations as represented within the WHOIS
database
across the RIRs, we
next turn to active prefix geolocation.
Our goal is thus to
help the community 
better understand where addresses are being used post-allocation,
whether registration information is accurate and can serve operational
needs, and inform ongoing discussion over RIR policy that are
important to maintaining the Internet's open and equitable future.

Figure~\ref{fig:method} provides an overview of our WHEREIS geo-audit
methodology, with three primary components: 
(1) ingesting RIR WHOIS into a database; (2) active
ICMP probing, which includes selecting the origin vantage point and 
destination targets
within prefixes; and (3) delay-based RIR consistency inferences. Important
supporting aspects of this high-level methodology include:

\begin{figure*}[t!] 
 \centering
 \resizebox{1.2\columnwidth}{!}{\includegraphics{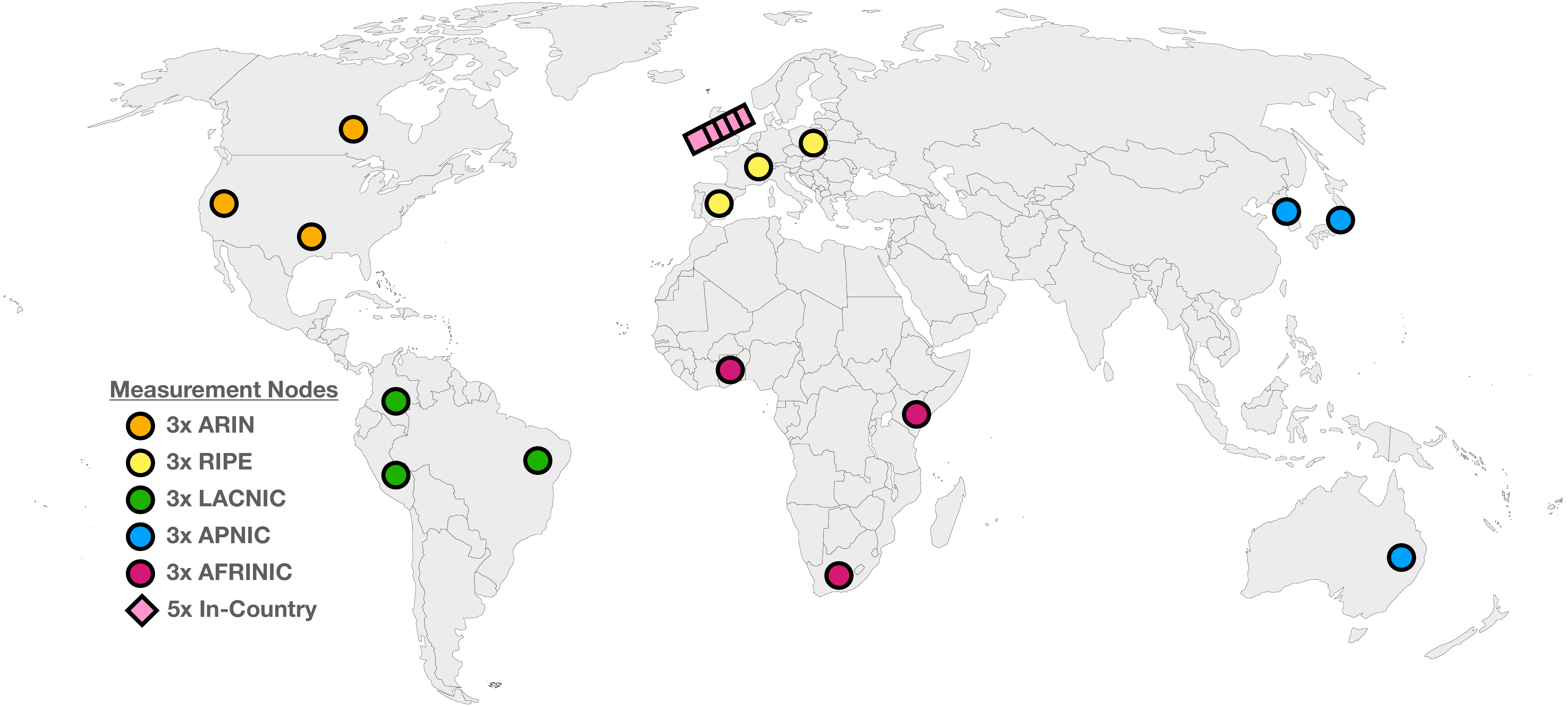}}
 \caption{WHEREIS uses RIPE Atlas and constraint-based geolocation 
  to infer true prefix locations.  To tighten the feasible 
  geolocation region, we use stable Atlas anchor nodes known to
  be in each RIR's region as well as nodes within the country 
  of the prefix's registered owner.  If the
  feasible geolocation matches \emph{either} the responsible RIR or the 
  RIR of the owner, we call the registration ``consistent,'' else
  ``inconsistent.''}
 \label{fig:atlas}
\end{figure*}

\begin{itemize}[leftmargin=*]
\punkt{Mapping countries to RIRs} 
We map countries 
and their corresponding ISO country codes,
to the responsible RIR.  Using publicly available information for each
registry, we map 244 different countries (including dependent
territories with their own ISO code) to the five RIRs using the
authoritative NRO mapping~\cite{nro-mapping}.
ARIN, RIPE, APNIC, LACNIC, and \afrinic are each currently responsible
for 29, 73, 54, 31, and 57 countries respectively.

\punkt{Target addresses} Delay-based active IP geolocation requires
a responsive address within a prefix.  Because
it focuses on stable IPv4 addresses, we utilize
the ISI Census IPv4 hitlist for this 
purpose~\cite{isihitlist}.
The hitlist includes a score for each 
address; to obtain address likely to respond, we filter the hitlist 
for addresses with a score $\ge99$.  We then 
longest-prefix map IPv4 hitlist addresses
to their respective RIR prefix
to generate candidate targets
within fine-grained RIR prefixes 
for active probing.

For IPv6, we use ICMPv6 responsive addresses from the IPv6 Hitlist Service \cite{gasser2018clusters,zirngibl2022rustyclusters,steger2023targetacquired}, specifically data from September 29, 2024, the latest available at the time of our experiments.
As aliased prefixes \cite{gasser2018clusters}---i.e., a single machine responding to all addresses in a possibly large prefix---are not present in the ICMPv6 responsive addresses, we investigate their prevalence in WHOIS data.
We find only 1\% of IPv6 WHOIS prefixes covered by the hitlist to be affected by aliased prefixes and thus conclude that using ICMPv6 responsive addresses without aliased prefixes are suitable targets for IPv6.

\punkt{Active Probing} The accuracy of constraint-based IP geolocation
is correlated with the distance from the target to the nearest vantage 
point; in our case, we desire vantages within the country where the prefix is
registered. For this purpose, we utilize the rich coverage afforded
by the RIPE Atlas project~\cite{bajpai2015survey}.  As of this writing, Atlas
has over 13k ``connected probes'' (active vantage points) and ``anchors'' 
(more stable and maintained nodes) with coverage in 
over 90\% of all
world countries~\cite{atlas}. 

\punkt{Vantage Location Filtering} While 
RIPE Atlas nodes include meta-data with their physical 
location, we follow best practices to ensure that the reported
locations are accurate.
We first filter out all suspected bad
probes in the list 
published by Izhikevich \etal~\cite{izhikevich2024trustverifyoperatorreportedgeolocation}.
Next, for the remaining alive and ``connected''
nodes, we follow the practice
of Darwich \etal~\cite{10.1145/3618257.3624801} and remove probes with bad meta-data,
including default country geo-coordinates. 

\punkt{Vantage Selection} From the remaining set of Atlas nodes
post-filtering, we
select a stable set of 10 vantage points within each RIR region. 
In addition, we select a stable set of 10 nodes per
country---these nodes ensure that we probe the target from 
\emph{within}
the same country as its registered organization.  
In selecting our vantage points, we choose anchors 
over probes, and prefer to select those in different
ASNs, in so far as possible (for some countries, there are fewer
than 10 anchors or probes).  
As shown in prior literature
as well~\cite{wang2011towards,10.1145/3618257.3624801}, the closest vantage has the most impact in
constraining the feasible geolocation space.  
The choice of 10 within-country
vantages and choosing those within different ASNs seeks to 
maximize the opportunity to find vantages near the target prefix
while adhering to our agreement to minimize load on the RIPE Atlas
platform, as well as its long-term measurement archival system.

\begin{figure*}[!t]
\centering
\begin{minipage}{0.60\textwidth}
\centering
{\small 
\resizebox{1.0\columnwidth}{!}{
\begin{tabular}{l p{4cm}|rrr} 
\toprule
                    &      & \multicolumn{3}{c}{Example} \\ 
Result              & Description & $RIR_{Reg}$  &  $RIR_{Org}$  & $RIR_{Geo}$ \\ 
\midrule 
($FC$) Fully Geo-consistent          &      
Geolocates in RIR and org's region
&ARIN &  ARIN    & ARIN \\ 
($OC$) Org Geo-consistent   &      
Geolocates outside RIR region, within org's region
&RIPE &  ARIN    & ARIN \\ 
($OI$) Org Geo-inconsistent &      
Geolocates in RIR's region, org is OOR
&ARIN &  RIPE    & ARIN \\ 
($RI$) Registry Geo-inconsistent     &      
RIR and org's region consistent, geolocates OOR
&ARIN &  ARIN    & RIPE \\ 
($FI$) Fully Geo-inconsistent        &      
RIR, org, and geolocation all in different regions
&ARIN &  RIPE    & APNIC  \\ 
\bottomrule
\end{tabular}}}
\end{minipage}
\begin{minipage}{0.37\textwidth}
\centering
\resizebox{1.0\columnwidth}{!}{\includegraphics{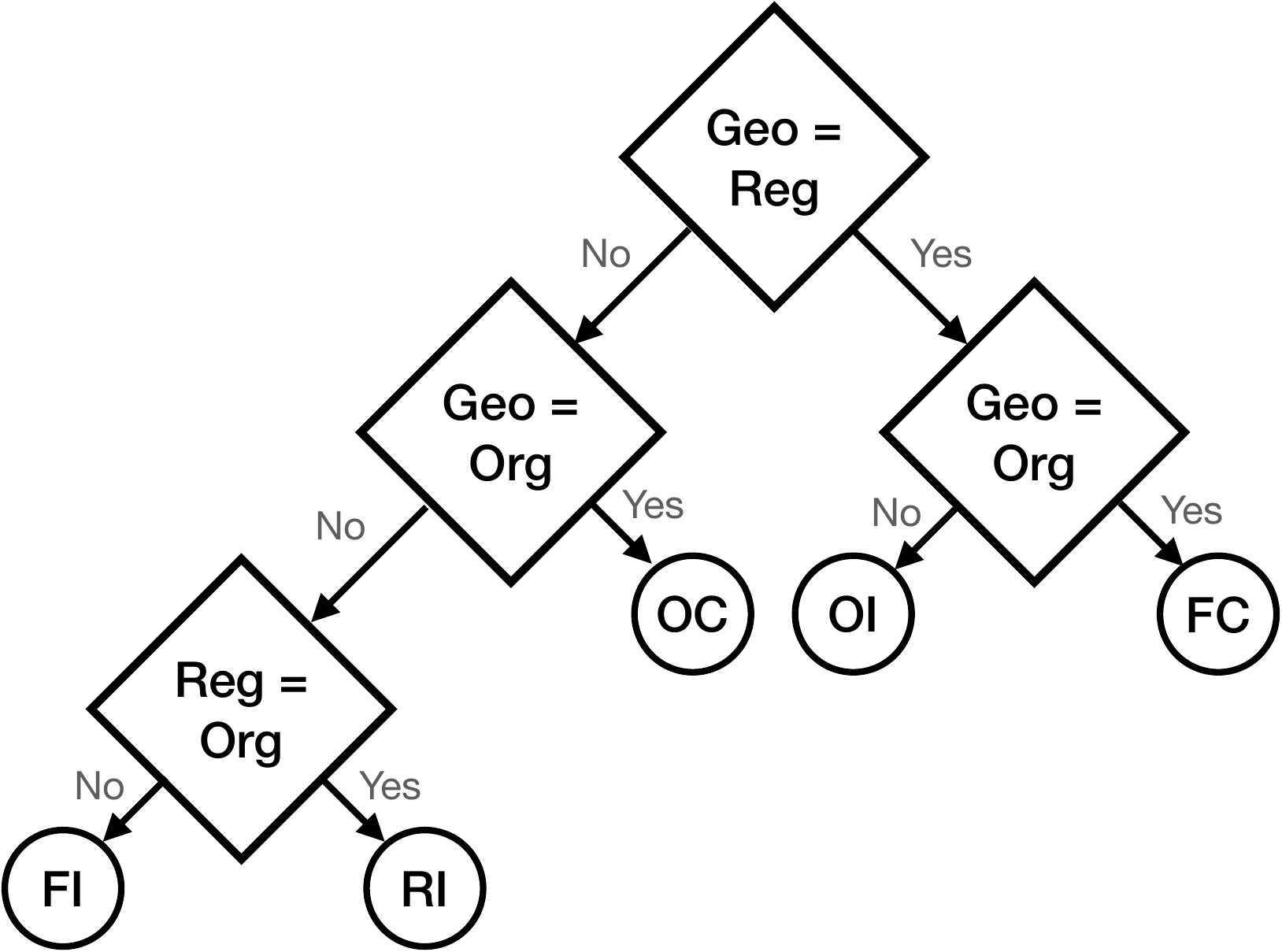}}
\end{minipage}
 \caption{Classification: Given a prefix allocated by
 $RIR_{Reg}$, with an organization in $RIR_{Org}$, we determine
 the RIR corresponding to the prefixes' inferred geolocation
 $RIR_{Geo}$ (\S\ref{sec:method:delay}). We then compare the three RIR's to 
 classify the prefix as either: 
(FC) Fully Geo-consistent; 
(OC) Organization Geo-consistent;
(OI) Organization Geo-inconsistent;
(RI) Registry Geo-inconsistent; or
(FI) Fully Geo-inconsistent.
 }
 \label{fig:terminology}
\end{figure*}

\punkt{Prefix handling} 
Central to this work are IP prefixes.
We construct a binary radix trie from the registered
prefixes and longest prefix match each hitlist target.  Thus if a target belongs to two
WHOIS prefixes, either in different RIR databases or within the same
RIR, we use that target to measure the consistency of the registration information
for the most specific prefix---this captures the inherent sub
delegation behavior common in prefix registrations.
When 
computing statistics, we take subnetting and
prefix aggregation into account so as to not 
double or over-count.

\punkt{BGP} We only make inferences for WHOIS prefixes that are either aligned
or subnets of the corresponding BGP advertisement (see \Cref{tab:bgp}) to avoid
mischaracterization due to traffic engineering or
content distribution.

\punkt{Anycast} A potential source of error in our geo-audit is 
anycast~\cite{sommese2020manycast2} wherein a prefix is advertised and reachable from multiple 
geographic regions.  We therefore utilize the current state-of-the-art
``AnyCatch'' technique and public service~\cite{anycatch}.
AnyCatch runs its own anycast instance, and issues ICMP probes
from different physical locations.  If the target is itself not
anycast, the ICMP responses should take the single shortest path back
to a single instance in the AnyCatch network.  However, if the target
is anycast, the responses will return to different nodes within
AnyCatch.  We ignore and do not geo-audit any prefixes that are
known to be anycast.

\end{itemize}

After identifying a likely responsive target within a prefix 
to geo-audit, we
instruct RIPE Atlas to issue ICMP measurements from 20 different vantage
points as depicted in Figure~\ref{fig:atlas}: 
three (3) vantages within the region of each of the five (5)  RIRs, and 
five (5) within the country to which the prefix is registered.  
All of the vantages are selected from the stable and filtered set of
vantages described previously.  Each
measurement is a one-off measurement, public, and tagged to
preserve the measurement artifact.  We then asynchronously fetch
results 
and insert them into our database.

\subsection{Speed-of-light-based RIR Inference}
\label{sec:method:delay}

As detailed previously in \Cref{sec:interregion}, the organization
to which a prefix is allocated may have a mailing address outside of
the RIR that manages the prefix's supernet.  For a prefix $P$ 
allocated to organization $Org$ by $RIR_{Reg}$, where the contact information
for $Org$ lists an address in country $CC$, we
define three important audit values: $RIR_{Reg}$ is the RIR in
which the prefix is registered; $RIR_{Org}$ is the RIR responsible
for $CC$, and $RIR_{Geo}$ is the RIR region where $P$ is located based on
delay-based geolocation.

From each of the 20 different vantages, RIPE Atlas performs three ICMP echo requests to
the target.  Among the responses (up to 60 ICMP echo replies) we find
the vantage returning the minimum RTT to the target to capture the
smallest propagation delay component.  As is well-established in the
literature, using the minimum RTT is a good proxy for
capturing propagation latency (which is fixed) and minimizing effects
that contribute variable delay (\eg queuing and processing delay).  
We then 
convert the round-trip latency to a distance based on the 2/3c
speed-of-light propagation in fiber optic cables
(\Cref{app:sol} examines the sensitivity of this parameter).  From the location of the
vantage, we find all physically possible countries and RIRs within that distance
radius, \ie those that do not violate the
speed-of-light propagation constraints.  
Note that while latency-based geolocation can
be inaccurate for precision geolocation, conversely it
provides a much higher degree of accuracy at the
country-level~\cite{2011-huffaker-gt}.  In
this work, we use latency-based geolocation at an even coarser
granularity---continents and RIR regions.  Further, we conservatively 
rely on
latencies to show that resources are \emph{not} in a particular RIR
region.

\subsection{Taxonomy of Prefix Registration Geo-Consistency}
\label{sec:method:taxonomy}

The value of $RIR_{Reg}$ and $RIR_{Org}$ can be one of the five
different RIRs, while $RIR_{Geo}$ (cf. \Cref{sec:method:delay}) is a 
set of possible RIRs we infer based on the 2/3c latency radius.
Given these values, we define five different possible inferences, which are summarized with an example in Figure~\ref{fig:terminology}.
A ``Fully Geo-consistent'' result is one where the registered RIR, RIR
of the organization's country code, and the RIR of the IP address geolocation determined by RIPE Atlas
all match.  A ``Organization Geo-consistent'' result is one where the
prefix's RIR is different than the RIR of the organization's mailing
address, but the inferred geolocation matches the organization's
country code RIR.  

The three remaining results show some form of unexpected
inconsistency. ``Organization Geo-inconsistent'' for instance occurs when
our inferred RIR geolocation matches the RIR of the prefix, but not
the organization's responsible RIR.  
Or, both the delegating RIR and the organization's responsible RIR
match, but the inferred geolocation differs from both, a
result we term ``Registry Geo-inconsistent.'' Finally, if all three
values are different, we term the result ``Fully Geo-inconsistent.'' 

\begin{table*}[t] 
\caption{Methodology validation against Speedtest server ground-truth
(91 servers).
Manual investigation of the Organizationally Consistent (OC) and
Registry Inconsistent (RI) prefixes confirms 100\% accuracy of the
methodology for these inferences.}
\label{tab:validate}
\resizebox{1.8\columnwidth}{!}{
\begin{tabular}{lrrrrr}
    \toprule
  &ARIN        &RIPE        &APNIC       &LACNIC      &AFRINIC\\
  \midrule
($FC$) Fully Geo-consistent &31 (86.1\%) &19 (86.4\%) &11 (84.6\%) &10 (90.9\%) &8 (88.9\%)\\
($OC$) Org Geo-consistent &1 (2.8\%)   &3 (13.6\%)  &2 (15.4\%)  &1 (9.1\%)   &0 (0.0\%)\\
($OI$) Org Geo-inconsistent &0 (0.0\%)   &0 (0.0\%)   &0 (0.0\%)   &0 (0.0\%)   &0 (0.0\%)\\
($RI$) Registry Geo-inconsistent&4 (11.1\%)  &0 (0.0\%)   &0 (0.0\%)   &0 (0.0\%)   &1 (11.1\%)\\
($FI$) Fully Geo-inconsistent&0 (0.0\%)   &0 (0.0\%)   &0 (0.0\%)   &0 (0.0\%)   &0 (0.0\%)\\
\bottomrule
\end{tabular}
}
\end{table*}

\subsection{Validation}
\label{sec:method:validation}

While we obtain and report on result validation obtained from RIRs and
providers in Sections~\ref{sec:results:case} and
\ref{sec:results:community}, we additionally validate our measurement
methodology. To do so, we utilize the IP addresses and ground-truth
geolocations of Speedtest servers~\cite{paul2022importance} within the different RIR
covered regions.  While these servers are intended for testing bulk
transfer capacity, they are well-connected, globally distributed,
responsive to our measurements, publicly published~\cite{speedtestsrv}, and
disjoint from the RIPE Atlas probes.  

We select 100 Speedtest servers as targets within the registered IP
prefix to which they belong. To find server addresses, we query the
Speedtest web API from five different network hosts around the world.
We obtain 36, 22, 13, 11, and 9 responsive servers with IP addresses
in ARIN, RIPE, APNIC, LACNIC, and AFRINIC registered space respectively
(91 total; 9 servers did not respond to ICMP requests).
For each of these servers, we utilize the published ground-truth
geolocations~\cite{speedtestsrv}.
We then use the entirety of our measurement and inference pipeline as
described in this Section on these targets and registered prefixes.

Table~\ref{tab:validate} displays our complete inferences.  Fully
Consistent (FC) inferences indicate that our methodology confirms
that the prefixes are connected in a region belonging to the 
RIR where the prefix is registered, and that the contact information
is also in the same RIR.  One ARIN, three RIPE, two APNIC, and one
LACNIC prefix are inferred to be 
Organizationally
Consistent (OC), and we manually verify the correctness of these
inferences by querying the WHOIS and comparing against the
ground-truth geolocation.  Finally, four prefixes registered in 
ARIN and one prefix registered in AFRINIC are Registry Inconsistent
(RI), indicating that WHEREIS geolocates the prefix outside of the 
registered RIR's region.  We again manually verify these against
the WHOIS and ground-truth locations.  In all five RI cases, we find
that these are correct inferences, and also that the registered
prefixes are subnetted within the BGP via announcements that are
more specific than the size of the WHOIS registration.  For instance, 
the AFRINIC RI inference is for a /20 prefix registered in AFRINIC
to an organization with an address in Burkina Faso within a /24 BGP
announcement.  However, WHEREIS
confirms, via constraint based latency, that the true location of
the target server in that prefix is in the USA.  The ground-truth
from Speedtest lists the server in Ashburn, VA.  Thus, WHEREIS made
a correct RI inference.

\subsection{Limitations}
\label{sec:method:limitations}

There are several potential limitations with our methodology.  First,
we may be unable to find an ICMP-responsive target within a particular
prefix.  Second, 
RIPE Atlas may not have any active probe within the country where a prefix
is registered, which might result in false negative results.
In these cases, we conservatively classify a prefix as geo-consistent, where in reality it is geo-inconsistent.
Our results, thus provider a \emph{lower bound} of geo-inconsistent prefixes.
Third, the reported
location of the Atlas node may be incorrect.  However, our use of
anchors and rigorous filtering, as described previously, attempts to mitigate any source of
error due to Atlas locations.
\section{Results}
\label{sec:results}

\begin{table}[t] 
  \caption{RIR prefix filtering pipeline.}
  \vspace{-4mm}
  \label{tab:filtering}
         \begin{tabular}{l r r}\hline
            \toprule
                                         & \textbf{IPv4} & \textbf{IPv6} \\
            \midrule
            Candidate RIR prefixes      & 94,755  & 56,838 \\
            IP targets probed           & 112,659 & 88,546 \\
            Responsive targets          & 82,253  & 85,884 \\
            Responsive prefixes         & 72,265  & 56,134 \\
             \hspace{4mm}- Anycast      & 51      & 118 \\
             \hspace{4mm}- NIR          & 3,669   & 7,728 \\
             \hspace{4mm}- BGP supernet & 6,455   & 5,203 \\
            \midrule
            Prefixes after filtering    & 62,090  & 43,085 \\
             \hspace{4mm}- Conflicting status & 6    & 85 \\
            \midrule
            Final Prefixes              & 62,084  & 43,000 \\
            \bottomrule
         \end{tabular}
  \vspace{-2mm}
\end{table}

We randomly select 20\% of the IPv4 prefixes and all IPv6 prefixes containing
at least one target in the hitlist from each of
the five RIRs, for geo-audit.  When a 
prefix contains more than one target, we choose two targets to
probe.  
In this fashion, we balance the 
objective of obtaining a representative sample while adhering
to our agreement with RIPE to not overburden the 
platform as described in~\Cref{sec:ethics}.

For transparency and reproducibility,
our measurement result sets are publicly available from RIPE Atlas
and can be found via the \anon{\texttt{neo-ip-20241018}}
tag.  Each
measurement's meta-data encodes the RIR prefix, date, and other
relevant information.

\subsection{Prefix Filtering}

This sampling yields 94,755 distinct candidate IPv4 prefixes and
112,659 targets (addresses within the prefix), and 56,838 distinct candidate IPv6 prefixes with 
88,546 IPv6 targets.  Of these targets, 82,253 IPv4 and 85,884 were
responsive to ICMP echo requests (\eg due to firewall filters or
an offline target in the hitlist), corresponding to 72,265 and 56,134
IPv4 and IPv6 responsive prefixes.

To focus on identifying non-ambiguous instances of
RIR policy violation, we filter
the candidate prefixes as summarized in  
Table~\ref{tab:filtering}.
First, as
described in~\S\ref{sec:method}, we remove 51 and 118 IPv4 and IPv6 anycast prefixes.  We
then remove 3,669 IPv4 and 7,728 IPv6 prefixes that 
belong to an National Internet Registry (NIR) as these may have policies
independent from the five RIRs.  Next, as described
in~\S\ref{sec:data:bgp}, we remove
6,455 IPv4 and 5,203 IPv6 prefixes that are either within a BGP 
supernet or cover mixed ASes.  Note that these numbers represent the
stepwise filtering operations and the order of operations impacts the
number of prefixes removed at each step as some prefixes belong to
multiple categories, \eg a prefix can be both anycast and within an
NIR.  

We probe targets within our
filtered set of prefixes over a 12 day period from October 9--21, 2024
with RIPE Atlas.
We find 6 IPv4 and 85 IPv6 prefixes where the two 
targets within the prefix produce conflicting geo-consistency
inferences.  We remove these conflicting status prefixes.  The
remainder of this section presents geo-consistency classification results for
the remaining 62,084 IPv4 and 43,000 prefixes according to the
method of \Cref{fig:method} and taxonomy of
\Cref{fig:terminology}.

\begin{table}[t] 
\caption{Overview of geo-consistency results.}
\label{tab:aggresults}
\vspace{-4mm}
\begin{tabular}{lrr} 
\toprule
Result                    & IPv4 Prefixes   & IPv6 Prefixes\\
\midrule 
Fully Geo-consistent      & 60,972 (98.2\%) & 42,019 (97.7\%) \\
Org Geo-consistent        & 653 (1.1\%)     & 739 (1.7\%) \\
Org Geo-inconsistent      & 255 (0.4\%)     & 128 (0.3\%) \\
Registry Geo-inconsistent & 166 (0.6\%)     & 106 (0.2\%) \\
Fully Geo-inconsistent    & 38 (0.1\%)      & 8 (0.0\%) \\
\bottomrule
\end{tabular}
\end{table}

\subsection{Prefix Geo-Consistency}
\label{sec:results:geo}

\begin{figure}[t!] 
 \centering
 \resizebox{1.0\columnwidth}{!}{\includegraphics{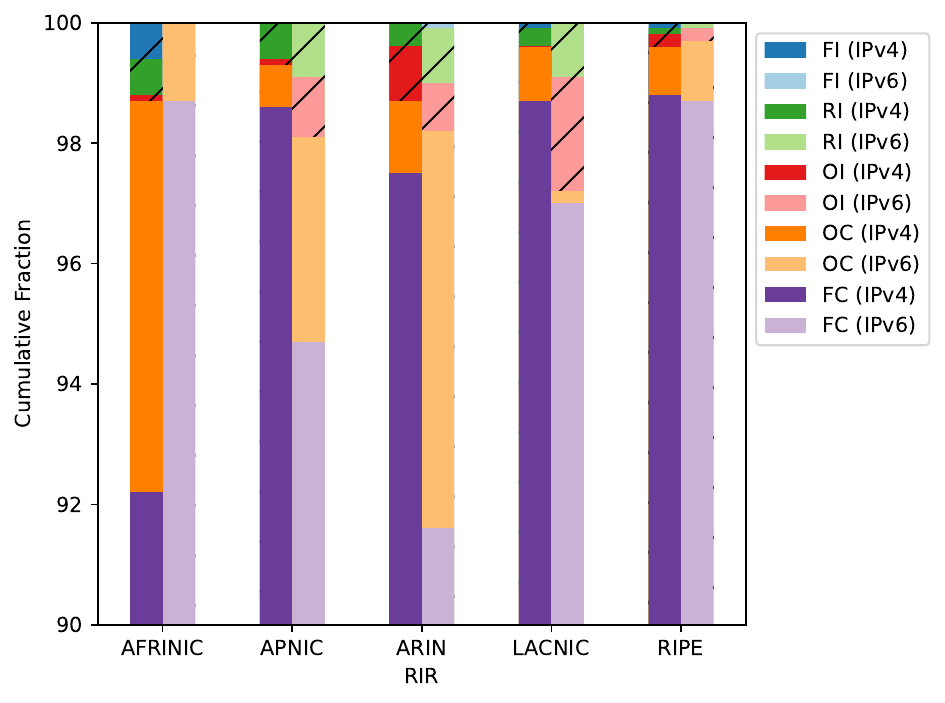}}
 \vspace{-5mm}
 \caption{Per-RIR WHEREIS results: >98\% of 
classified prefixes are fully or organization consistent (FC/OC).
}
 \label{fig:distrib}
 \vspace{-4mm}
\end{figure}

Table~\ref{tab:aggresults} presents the aggregate geo-audit results,
while Figure~\ref{fig:distrib} shows the per-RIR classification
distribution.
We observe that, in aggregate and as a relative proportion of
the prefixes audited, the registry information is largely
consistent: overall 98.2\% of IPv4 prefixes are fully 
geo-consistent.  RIPE exhibits the most
consistency (over 99.6\%).
APNIC and LACNIC are also highly consistent, while ARIN is
mostly consistent (98.8\%).
Furthermore, \afrinic exhibits markedly less geo-consistency
with only 92.2\% of prefixes being fully consistent.
\afrinic also leads in fully geo-inconsistent
prefixes (nearly 1\%).

In IPv6, we find overall a higher share of prefixes being consistent, with 98.9\% being either fully geo-consistent or organization geo-consistent.
Similarly as in IPv4, the majority of inconsistencies are either
organization or registry, with low fractions of fully geo-inconsistent prefixes.
When investigating consistency levels across RIRs in IPv6, we see quite a number of similarities but also differences compared to IPv4.
For similarities, we see that RIPE and  APNIC are also highly geo-consistent in IPv6.
On the contrary, AFRINIC is highly geo-consistent in IPv6, whereas
ARIN exhibits the lowest fraction of geo-consistent IPv6  prefixes.

\begin{figure*}[t] 
 \centering
\begin{subfigure}{0.5\textwidth}
  \centering
  \includegraphics[width=.9\linewidth]{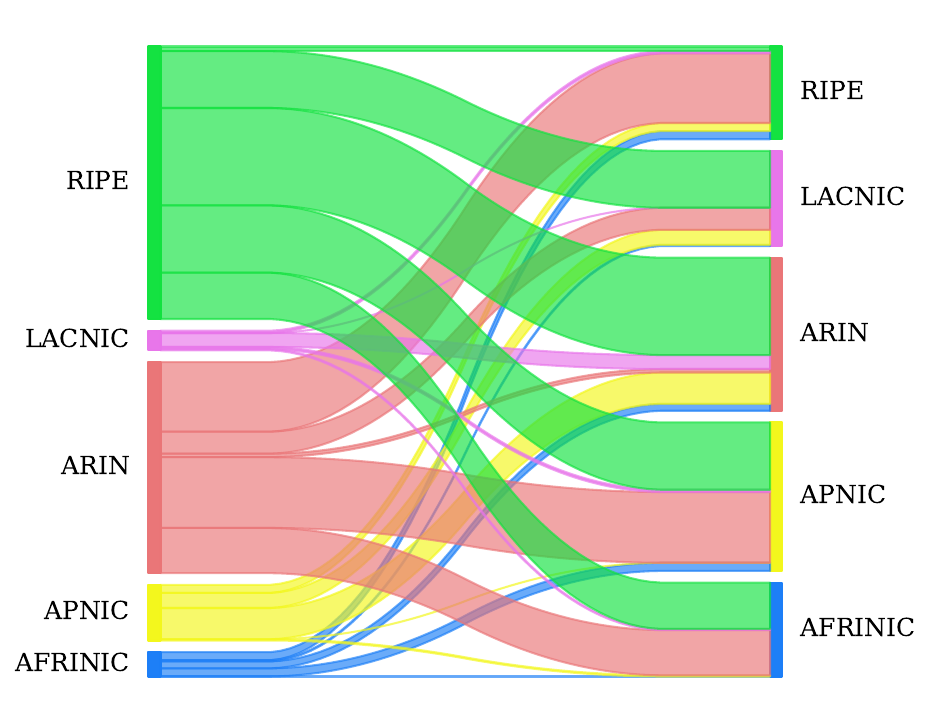}
  \caption{IPv4.}
  \label{fig:girp4}
\end{subfigure}%
\begin{subfigure}{0.5\textwidth}
  \centering
  \includegraphics[width=.9\linewidth]{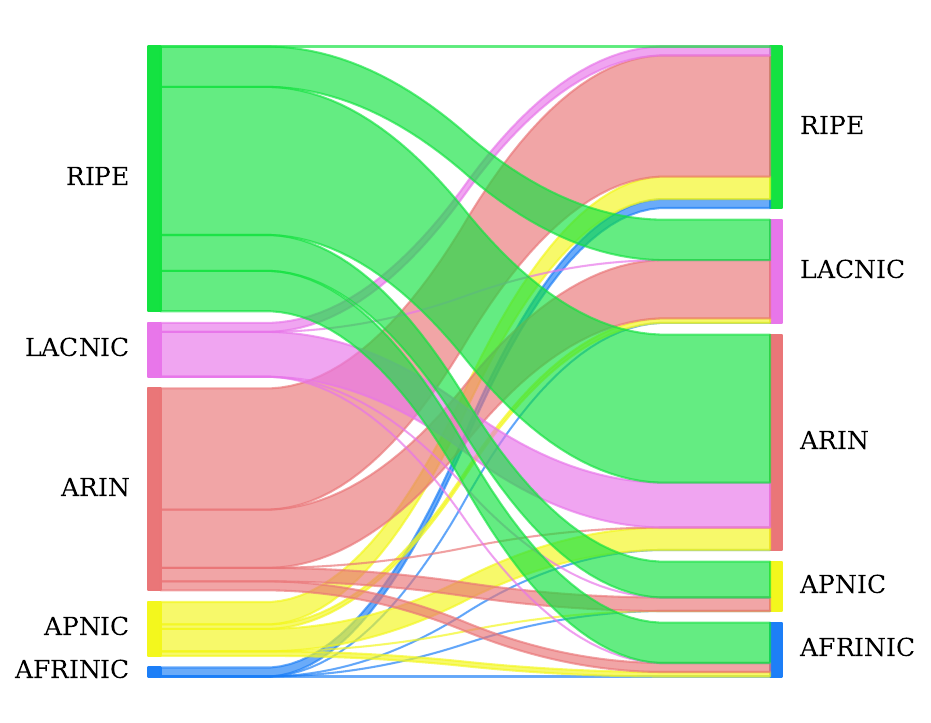}
  \caption{IPv6.}
  \label{fig:girp6}
\end{subfigure}%
 \caption{Geo-inconsistent prefixes in IPv4 (left) and IPv6 (right). 
Source of Sankey is the inconsistent prefix's RIR, Sankey destination 
is the WHEREIS inferred RIR region.}
 \label{fig:geoir}
\end{figure*}

Delving into the IPv4 inconsistencies (see \Cref{fig:girp4}), we find that the primary contributor
(approximately 75\%) to ARIN geo-inconsistencies are prefixes that our
audit shows to be in the APNIC or RIPE region.  Over 50\% of the LACNIC
geo-inconsistencies are prefixes that are physically in the United
States.  The geolocation of geo-inconsistent \afrinic prefixes is
dominated by Asia, but also with an appreciable number
of prefixes that geolocate to ARIN and RIPE countries.
In IPv6 (see \Cref{fig:girp6}), almost one fourth of ARIN geo-inconsistent prefixes is located in Switzerland, with the remainder being found in countries in the LACNIC (\eg Brazil, Mexico) and RIPE region (\eg France, the Netherlands, Germany).
Similarly as in IPv4, the majority of inconsistent LACNIC prefixes are located in the ARIN region (US and Canada).
Finally, RIPE geo-inconsistent IPv6 prefixes are also dominantly in the ARIN region.

\subsection{Prefix Characteristics}
\label{sec:results:properties}

\begin{figure*}
  \centering
  \begin{subfigure}{0.48\textwidth}
    \centering
    \includegraphics[width=0.9\linewidth]{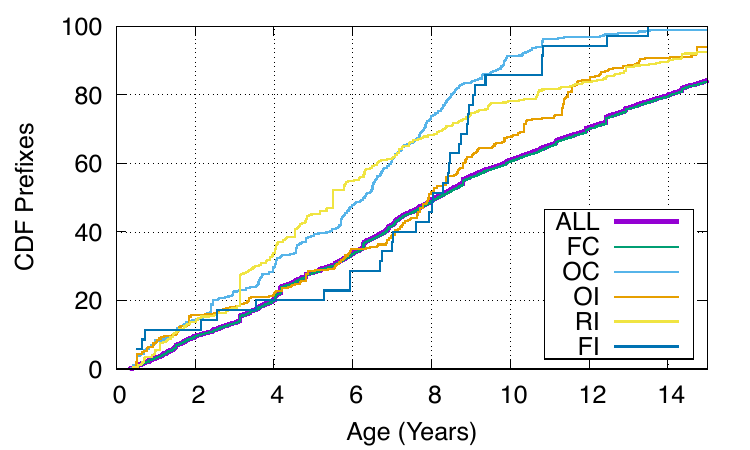}
    \caption{IPv4 last updated}
    \label{fig:properties:v4age}
   \end{subfigure}
  \begin{subfigure}{0.48\textwidth}
    \centering
    \includegraphics[width=0.9\linewidth]{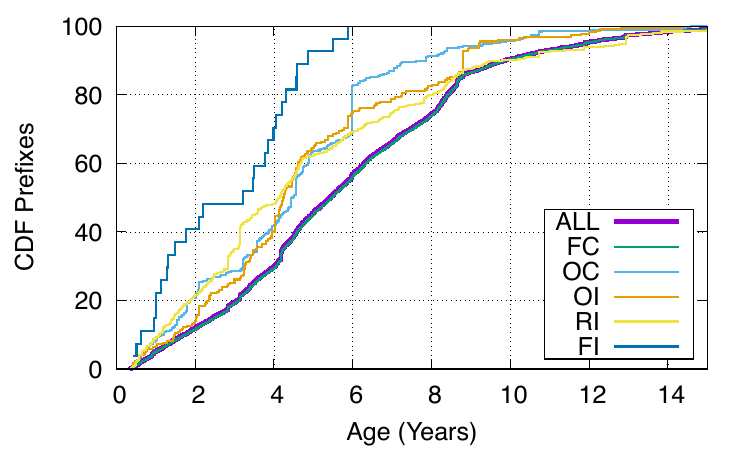}
    \caption{IPv6 last updated}
    \label{fig:properties:v6age}
   \end{subfigure}
    \caption{Relationship of geo-consistency to prefix characteristics.}
    \label{fig:properties}
\end{figure*}

First, we examine the relationship of a prefix's age, length and
allocation type to its inferred consistency.  WHOIS records frequently
contain timestamps for when the prefix was first registered as well as
last updated. Because not all RIRs provide the first registered time,
and because we are most interested in the recency of records, we use the
last updated field as a measure of the prefix's age.
\Cref{fig:properties:v4age,fig:properties:v6age} show
the distribution of IPv4 and IPv6 prefix last-updated age versus the
inferred consistency.  In general, and counter-intuitively, prefixes
that are more recently updated are more likely to exhibit an
inconsistency, although this correlation is more pronounced in IPv6.  

\begin{figure}
    \includegraphics[width=0.9\linewidth]{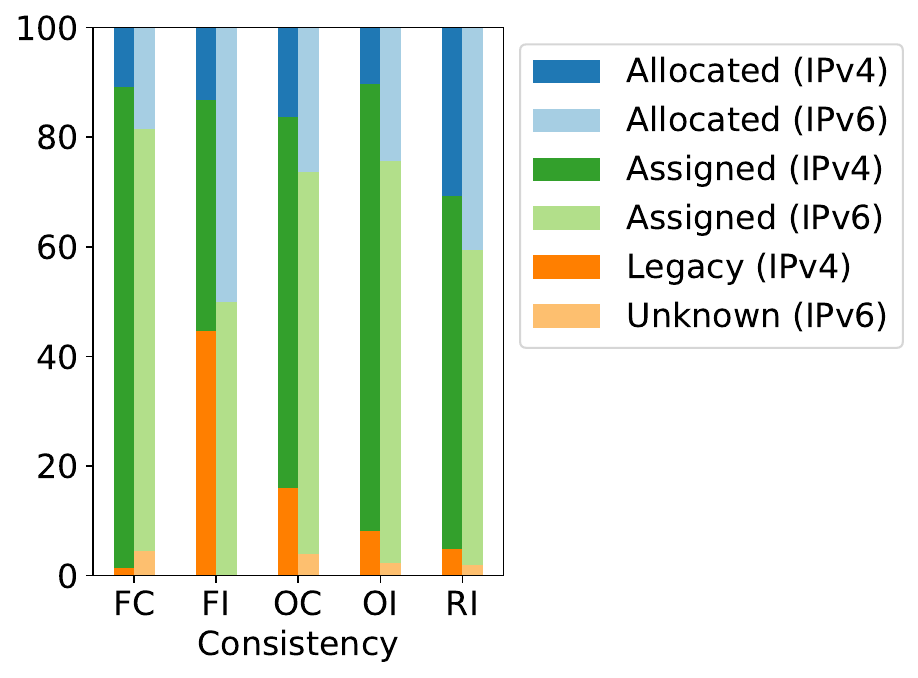}
    \vspace{-3mm}
    \caption{Geo-consistency across prefix allocation type.}
    \label{fig:properties:nettype}
  \centering
\end{figure}

Next, we correlate the allocation type to inconsistency results.
\Cref{fig:properties:nettype} shows the distribution of
allocated, assigned, and legacy (IPv4) or unknown (IPv6) prefixes for each inferred
consistency category across IPv4 and IPv6.  The majority of fully
consistent IPv4 prefixes are assigned, wherein the prefixes are
intended for use by the recipient organization.  For IPv6, most
inconsistencies are attributable to allocated prefixes, or those
intended for subsequent distribution by the recipient organization.
Legacy prefixes constitute a substantial
fraction of fully inconsistent prefixes, suggesting that the
registered information for these legacy assignments is not up-to-date.
As legacy addresses were allocated by IANA and predate the RIR system, they do not need to adhere to current RIR policies, this
is not wholly unsurprising, but still poses challenges to the 
application and uses of legacy registry data.

\section{Impact}
\label{sec:impact}

In this section we provide details on how geographically inconsistent WHOIS prefixes can impact Internet business models such as IP address leasing, legal proceedings regarding out-of-region transfers for AFRINIC, incorrect geolocation mappings present in popular geolocation databases, the impact of the used measurement platform, and engagement with the wider Internet community.

\subsection{IP Address Leasing}
\label{sec:impact:leasing}

Separate from IP address transfers, the scarcity of address space has 
driven a secondary market for leasing IPv4 addresses.  
Du~\etal~\cite{10.1145/3646547.3689010} recently developed a
methodology to infer IPv4 leasing and found 4.1\% of all IPv4 
address space to be leased.   
We obtain and use Du \etal's data to determine the extent to which our
inferred geo-inconsistencies are potentially explained by leasing.

Among the prefixes that we infer to be fully inconsistent, we 
find none of Du \etal's leased prefixes present for RIPE, LACNIC, or APNIC.
12.5\% of the fully inconsistent AFRINIC prefixes and 
27.3\% of the fully inconsistent ARIN prefixes are inferred to be
leased prefixes.  For registry inconsistent prefixes, we find 
0.0\%, 11.1\%, 12.7\%, 19.0\%, and 42.9\% overlap with the leased
prefix dataset for LACNIC, APNIC, ARIN, RIPE, and AFRINIC
respectively.

These findings suggest that IP leasing is a contributing factor to 
geo-inconsistent prefixes, but cannot explain the majority of our
inferences.
We reach out to IPXO, one prominent IP leasing platform provider present in our data and discuss these findings with them (see \Cref{sec:results:community}).

\subsection{AFRINIC Case Study}
\label{sec:results:case}

AFRINIC policy stipulates that resources are for the AFRINIC service
region, and any use outside the region should be solely in support of
connectivity back to the AFRINIC region \cite{afrinic_policy}
(see~\Cref{tab:nro}) . In view of the current litigation ~\cite{afrinic-court,afrinic-faq,ci-seized,sa-heist,krebs,crumbs} and discussions around out-of-region usage in the \afrinic region, we perform a complete WHEREIS 
examination of all \afrinic IPv4 prefixes for which there is a responsive
target in our hitlist; in total we are able to classify 7,208 \afrinic
registered prefixes in December, 2024.

Among these prefixes, we find approximately 91\% are fully consistent,
while approximately 2\% are either registry inconsistent or fully
inconsistent.
Comparing with our randomly sampled measurements in
\Cref{sec:results:geo}, we find a higher fraction of inconsistent
prefixes in AFRINIC.
These inconsistent prefixes are part of 51 different  
prefixes advertised in the global BGP and span 23 different autonomous
systems.  One BGP prefix accounts for approximately 17\% of all
inconsistencies, and two ASNs account for more than 53\% of the
inconsistencies we discover.  We omit naming these specific ASN, as 
one of the stipulations of our interaction with the RIRs was to 
only report aggregate results and not identify individual ASNs
found to be inconsistent.

Due to the timeliness of our work, we share these inconsistencies with
representatives of \afrinic.  The two largest contributing ASNs were
known to \afrinic and they validated our findings.  Out of a sample of
106 prefixes shared, 69 were found Fully Geo-Inconsistent (FI) and 37
Registry Inconsistent (RI).  AFRINIC staff noted that 19 prefixes were
known and of accepted use (Anycast, Global backbone nodes, peering
nodes, VSAT, etc), 77 prefixes were known and of unaccepted use as per
\afrinic policies, while 10 prefixes were unknown to them bearing
characteristics of both acceptable and non-acceptable use.
 \textbf{Note that we did not 
specifically target the measurement of any network or prefix, rather these
inconsistencies were identified as part of our complete audit of
\afrinic address space---suggesting that WHEREIS correctly
identifies geo-inconsistencies}.

\subsection{Influence on Geolocation Databases}
\label{sec:impact:geodb}

\begin{figure}[t!] 
 \centering
 \resizebox{0.9\columnwidth}{!}{\includegraphics{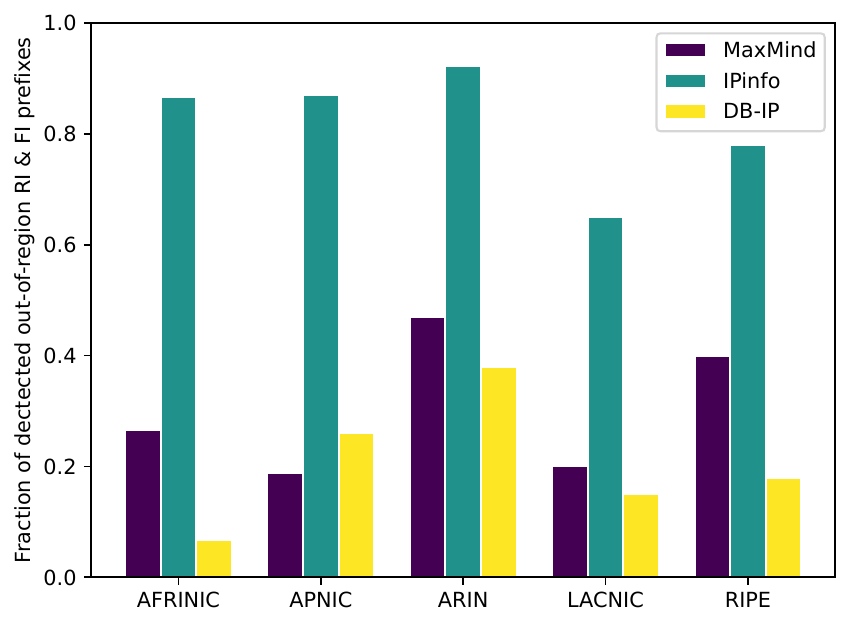}}
 \vspace{-2mm}
 \caption{Fraction of correctly detected out-of-region registration inconsistent (RI) and fully inconsistent (FI) prefixes by RIR and geolocation provider.
     IPinfo has the highest fraction of detected OOR prefixes, followed by MaxMind and DB-IP.
 }
 \label{fig:geoloc}
 \vspace{-1mm}
\end{figure}

Next, we evaluate the possible impact of geo-inconsistent RIR prefixes on geolocation databases.
To this end, we map registration inconsistent (RI) and fully inconsistent (FI) prefixes using three popular commercial geolocation providers: MaxMind \cite{maxmind}, IPinfo \cite{ipinfo}, and DB-IP \cite{dbip}.
As shown in \Cref{fig:geoloc}, we find large differences between geolocation providers for identifying out-of-region (OOR) prefixes with detection fractions ranging from 7\% to 92\% depending on geolocation database provider and RIR.
IPinfo has the highest share of detected OOR prefixes, ranging from 65\% to 92\%, while DB-IP has the lowest share with 7\%--38\%.
Looking at per-RIR detection ratios, we find that IPinfo detects the most OOR prefixes across all five RIRs, with MaxMind being in second place in AFRINIC, ARIN, LACNIC, and RIPE, while DB-IP performs second-best in APNIC.
Next, we investigate, in which region each geolocation provider does best:
All three geolocation providers have the highest detection ratio for ARIN prefixes with 47\% for MaxMind, 92\% for IPinfo, and 38\% for DB-IP, respectively.
On the other side of the spectrum, MaxMind has the lowest percentage of detected OOR prefixes for APNIC (19\%), IPinfo for LACNIC (65\%), and DB-IP for AFRINIC (7\%).
This shows that IPinfo is able to detect the large majority of OOR prefixes, while two other prominent geolocation providers might still be relying on location information in WHOIS data, which might not necessarily reflect the actual used location of IP addresses.

\subsection{Impact of Measurement Platform}
\label{sec:impact:platform}

Next, we analyze the impact of the used measurement platform on the detection of geo-inconsistent prefixes.
We coordinate with a commercial measurement platform provider and compare our results obtained from RIPE Atlas measurements with results using their measurement platform\footnote{Their measurement network is distributed across more than 800 servers, located in 370 cities and 120 countries.}.
We instruct them to use the exact same WHEREIS methodology, with the only difference being to substitute RIPE Atlas with their measurement platform and the selected targets to determine geolocation inconsistent prefixes.
As the measurement platform provider runs large-scale measurements towards many more targets compared to our RIPE Atlas measurements, they are able to detect 45.7k registry or fully inconsistent IPv4 and 1.5k IPv6 prefixes, compared to our 204 detected IPv4 and 114 detected IPv6 prefixes.

For prefixes audited by both RIPE Atlas
and the measurement platform provider, the commercial provider
has better detection recall than Atlas due to 
the larger scale of measurements and 
ability to leverage probe vantages closer to the targets.
Out of the 82.0k jointly measured IPv4 prefixes, 777 are detected as registration or fully inconsistent.
Of these detected IPv4 prefixes, 25.6\% are detected by both, 0.3\% by RIPE Atlas only, and 74.1\% only by the commercial measurement platform.
For the 55.5k jointly measured IPv6 prefixes, 827 are classified as
inconsistent.  In IPv6, the picture looks similar, with 11.5\% both
detected, 0.7\% by RIPE Atlas only, and 87.8\% by the commercial provider
only.

While the hitlist of responsive targets and limits of the measurement
platform may only provide a lower bound of detected OOR prefixes, RIPE
Atlas remains a suitable public platform to detect geo-inconsistent
WHOIS prefixes and has the benefits of reproducibility and repeatability.

\subsection{Community Coordination}
\label{sec:results:community}

We attempted to solicit validation and additional information
from network points of contact as registered in the WHOIS.
Surprisingly, there is substantial variation in the fraction of
prefixes that have a registered technical, abuse, or administrative
point of contact with an email address between RIRs.
Table~\ref{tab:poc} shows that while 99\% of the prefixes within
ARIN have at least one registered email point of contact, only
41\% of RIPE IPv4 prefixes and 26\% of RIPE IPv6 prefixes do.  
Similarly, fewer than 30\% of AFRINIC's prefix registrations contain
email addresses.
Furthermore, in January 2024, AWS announced that they plan to discontinue WHOIS lookups to retrieve email addresses due to their low success rate \cite{aws}.

Despite this, we make a best effort attempt to contact individual 
networks.  We aggregate results by email point of contact, and find
instances where there are both consistent and inconsistent prefixes
belonging to the same point of contact.  In this way, our hope is
to understand cases where administrators choose a particular 
registry or policy for different portions of their network.  We 
sent our inferences and results to 16
different network points
of contact. Unfortunately, at the time of writing we received no responses.

Throughout this work, engaged with the community to share
results and incorporate feedback.  In particular, we shared our
results with three of the five RIRs, ARIN, RIPE, and AFRINIC.  Our
interaction with AFRINIC is detailed in~\S\ref{sec:results:case}.

ARIN's main feedback was that our identified fully inconsistent
prefixes were indeed correctly identified by our methodology, and that
they primarily represented ``hosted services.''  In particular, ARIN
called out an instance of ``a hosting provider using ARIN IP addresses
for a datacenter in Tokyo, whose end user is located in Estonia.''  In
other instances, especially where ARIN addresses are in use in Mexico,
ARIN was not able to add significant amplifying detail, but
hypothesized that the service is being delivered in Southern
California or Texas, or that resellers are providing service across
the border.  ARIN stressed that any mismatch between the registry
information and the service delivery location is not a violation of
their policy.

Furthermore, as we find that a non-negligible share of our identified inconsistent prefixes are leased prefixes (see \Cref{sec:impact:leasing}).
We engaged with one of the largest IP leasing platform providers,
IPXO.  We shared 10 detected inconsistent AFRINIC prefixes with IPXO for validation and feedback.
Of the 10 AFRINIC prefixes our methodology determines five being located in the RIPE NCC region and the other five in the APNIC region.
Upon checking with their customers, IPXO updated the WHOIS entry of two prefixes to reflect the change in location that we detected.
For another two prefixes the customer disagreed with our assessment, even as we provided additional supporting evidence consisting of traceroute data and additional latency measurements to substantiate our inconsistency findings.
For the remaining six prefixes IPXO was unable to get confirmation by the customers regarding location changes.
This shows that personal contacts and engagement with the wider Internet community, an IP leasing platform provider in this case, can have measurable impact by correcting WHOIS entries, although it remains difficult to do this at scale.

\begin{table}
\caption{Fraction of prefixes with registered email point of contact (POC).}
\label{tab:poc}
 \begin{tabular}{lcc}
\toprule
         & \multicolumn{2}{c}{Prefixes with Email POC (\%)} \\ 
  RIR    & IPv4 & IPv6 \\ 
\midrule 
  ARIN   & 99.0 & 99.2 \\
  RIPE   & 41.4 & 26.3 \\
  APNIC  & 99.1 & 98.4 \\
  LACNIC & 94.9 & 91.0 \\
  AFRINIC& 29.1 & 23.5 \\
\bottomrule
\end{tabular}
\vspace{-3mm}
\end{table}

\section{Related Work}
\label{sec:related}

While the present work focuses on IPv4 and IPv6 addresses, the WHOIS 
protocol is also used to provide access to domain name and autonomous
system registration data.
Prior work shows that domain names are frequently used for abusive or
malicious purposes and registration behavior can be indicative of such
misuse~\cite{leontiadis2014empirically}.  Lauinger \etal analyze the
re-registration of domain names after their expiration and show how
attackers can leverage residual trust by capturing these expired
domains~\cite{lauinger2016whois}.  More recently, Lu \etal dive into
WHOIS in the era of GDPR to better understand domain registration
privacy~\cite{lu2021whois}.  
Nemmi \etal examined autonomous system registrations in the WHOIS, and
compared registrations against BGP announcements over
time~\cite{nemmi2021parallel}.  
We similarly examine the
correspondence between WHOIS and BGP, but at the addressing
granularity.
Although these prior works all utilize
WHOIS registration data, they do not consider 
IP address block allocations.

Hsu \etal \cite{hsu2023fiat} analyzes IPv6 prefixes in different data
sources, including WHOIS databases.
They find that the most common prefix lengths in WHOIS databases are /64, /48, and /56---with substantial differences between RIRs, with /64 prefixes prevalent in RIPE compared to the dominance of /48 prefixes in APNIC and AFRINIC.
However, Hsu \etal do not investigate the location information in the
WHOIS databases or the geolocation of prefixes.

Dainotti \etal \cite{dainotti2016lost} analyze both passive and active data, including
WHOIS registrations and BGP routing, to better understand the
utilization of the IPv4 address space.  This effort used a commercial
geolocation service to locate each /24, and showed the state of IPv4
utilization at the RIR and country levels in 2013.  Our work instead
explicitly considers current WHOIS georegistration inconsistencies, for both
IPv4 and IPv6.  

The scarcity of IP addresses has created markets where addresses
can be transferred and sold as a commodity~\cite{prehn2020wells}.
Livadariu \etal analyzed
transfers published by the RIRs to characterize the size of prefixes
and their eventual use as evidenced in the global BGP
table~\cite{2013-livadariu-fltm, 2017-livadariu-itmw}.  
In contrast,
we focus on understanding the true location where addresses are
used, and whether these locations are within or outside of the
regions for which the corresponding registry is responsible.  

More recently, Du \etal design a methodology to examine WHOIS
prefix allocations and their corresponding BGP AS origins to 
infer subnet leasing~\cite{10.1145/3646547.3689010}.  While Du \etal's
work does not examine geolocation, we analyze the extent to which
the inconsistencies we discover can potentially be explained by
leasing in~\Cref{sec:impact:leasing}.
While
unofficial ``under-the-table'' transfers where the correct location
and registry information is not properly updated may explain some
of the location inconsistencies we discover, we leave causal 
analysis to future work.

Many research efforts in the past two decades consider
IP address geolocation
\cite{katz2006towards,poese2011ip,komosny2017location,shavitt2011geolocation,wang2011towards,gharaibeh2017look,endo2010whois,scheitle2017hloc,du2020ripe,gamero2025empirically}.
These studies mainly focus on the accuracy and techniques for IP geolocation.
In contrast, we leverage IP address geolocation to infer
inconsistencies of entries in WHOIS databases and their actual
physical location.  These prior works further highlight challenges in using
speed-light constraints for geolocation.  While the goal of our paper
is not to develop a new or better delay-based geolocation method, we
note these prior findings to both analyze the sensitivity to the
speed-of-light constraints in our WHEREIS system, as well as guide our
section of a conservative inference methodology that prioritizes precision over recall.

More closely related to our work, Zander examined the accuracy of IP
allocations against geolocation databases~\cite{zander2012accuracy} 
and found a roughly 5\% difference by address space.
However, this effort, which is over a decade old, assumes the geolocation database as ground truth,
as assumption we cannot make as many geolocation databases are
themselves constructed in part from WHOIS; we examine the effect
of geo-inconsistencies on three commercial geolocation databases
in~\S\ref{sec:impact:geodb}.  
To the best of our knowledge, our work is the first to systematically 
analyze geo-inconsistencies in the WHOIS.

\section{Conclusions}
\label{sec:conclusions}

This work took the first steps towards understanding the geographic allocation and use of RIR prefixes.
We defined a taxonomy to characterize different levels of geo-inconsistencies.
Using RIPE Atlas we ran a measurement campaign to infer the geographical region of prefixes and identified 98\% as being geo-consistent.
We coordinated with three different RIRs with whom we exchanged data for further validation and discussion.
We hope these results can meaningfully contribute to the important
policy and technical discussions surrounding IP address allocation, as
well as highlight areas where RIRs may wish to work with their
constituent membership to improve registration accuracy.

\begin{acks}
We thank RIPE for maintaining the Atlas measurement infrastructure.
John Curran, Thomas Krenc, Anita Nikolich, Eric Rye, and John Sweeting provided
invaluable early feedback. We express our gratitude to AFRINIC staff for validating our findings for the case study and to Ben Du for providing the IP
leasing data.
\end{acks}

\label{page:end_of_main_body}

\balance
\bibliographystyle{ACM-Reference-Format}
\bibliography{neo}

\appendix
\section{Ethics}
\label{sec:ethics}

We applied for access to ARIN's bulk WHOIS data under their research
use provision and abide by ARIN's acceptable use policy.  The
remaining four WHOIS databases are publicly available.  Our work does
not involve human subjects or personally identifiable information.  Our
measurements use RIPE Atlas to send relatively innocuous ICMP probes
and follows best practices for active Internet measurements.
Furthermore, we contact the RIPE Atlas team to discuss our measurement approach with them.
To reduce the load on the RIPE Atlas system, the RIPE NCC suggested that we measure 20\% of prefixes for IPv4 and all prefixes in IPv6.
Finally, we respect RIPE Atlas resource limits and perform exponential back-off in case the RIPE Atlas API returns an error.
As such, this paper does not raise ethical issues.

\ifx\false
\section{Algorithm}
\label{app:algorithm}

We provide pseudo-code for our geo-audit inference algorithm here for
additional detail.  The algorithm assumes a database operation,
db\_lookup, (as described in our methodology) that performs a
longest-prefix match on a prefix $P$ to return the RIR to which it is
registered.  It further assumes (again, as described in our
methodology) a lookup function, rir\_lookup,  between an ISO country
code and the responsible RIR.  Finally, we abstract the RIPE Atlas
probing logic and assume a function that returns the set of ICMP
RTTs as produced by our active measurements in the atlas\_results
function.

\begin{algorithm}[]
\caption{GeoAudit($P$)}
\label{alg:audit}
\begin{algorithmic}
\If {$P\in anycatch$}
  \State return($AC$)
\EndIf
\State $RIR_{Reg}\gets$ db\_lookup($P$)
\State $CC\gets$ db\_lookup($P$)
\State $RIR_{Org}\gets$ rir\_lookup($CC$)
\State $RTT[]\gets$ atlas\_results($target \in P$)
\State $Probe_{min} = \underset{k}{\mathrm{arg\_min}}$ $RTT[k]$ 
       $\forall k\in probes$
\State $RIR_{Geo}\gets$ rir\_lookup($Probe_{Min}$)
\If {$RIR_{Geo} = RIR_{Reg}$}
  \If {$RIR_{Geo} = RIR_{Org}$}
    \State return($FC$)
  \Else
    \State return($OI$)
  \EndIf
\Else
  \If {$RIR_{Geo} = RIR_{Org}$}
    \State return($OC$)
  \ElsIf {$RIR_{Reg} = RIR_{Org}$}
    \State return($RI$)
  \Else
    \State return($FI$)
  \EndIf
\EndIf
\end{algorithmic}
\end{algorithm}

The algorithm returns one of six possible results: anycast ($AC$), 
Fully Geo-consistent ($FC$), Organization Geo-consistent ($OC$), 
Organization Geo-inconsistent ($OI$), Registry Geo-inconsistent ($RI$)
or Fully Geo-inconsistent ($FI$).  Section~\ref{sec:method} details
each step of our audit methodology, while Section~\ref{sec:results}
provides results from a 50k prefix RIR geo-audit.

\fi

\ifx\false
\section{WHOIS Analysis}

To better understand the relationship of OOR prefixes between RIRs, we next analyze from which to which RIR these out-of-region prefixes exist.
\Cref{fig:ir} shows a Sankey diagram of the inter-region
registration activity.  Most of the out-of-region \afrinic
IPv4 registrations are from organizations in Asia and America.
Interestingly, approximately the same number of ARIN IPv4 prefixes are 
allocated to European organizations as RIPE prefixes are allocated 
to American organizations.  
Given the historical nature of ARIN, we see the majority of inter-region IPv4 prefixes allocated by ARIN and used by LACNIC, APNIC, and RIPE.
In IPv6---shown in \Cref{fig:irp6}---the picture looks quite different with RIPE being the largest source of out-of-region WHOIS prefixes, which are mostly used by organizations registered in the ARIN and LACNIC regions.
Again, we see a similar number of ARIN prefixes used by RIPE region organizations as the other way around.
Contrary to IPv4, \afrinic plays almost no role in out-of-region registrations in IPv6, again hinting at the focus on valuable IPv4 address space.

\begin{figure*}[t] 
 \centering
\begin{subfigure}{0.5\textwidth}
  \centering
  \includegraphics[width=.9\linewidth]{rir.pdf}
  \caption{IPv4}
  \label{fig:irp4}
\end{subfigure}%
\begin{subfigure}{0.5\textwidth}
  \centering
  \includegraphics[width=.9\linewidth]{rir_v6.pdf}
  \caption{IPv6}
  \label{fig:irp6}
\end{subfigure}%
 \caption{Inter-region WHOIS registration: proportions of IPv4 (left)
and IPv6 (right) prefixes allocated by an RIR (source) to an organization with a
physical address in a region for which a different RIR (destination) is
responsible.}
 \label{fig:ir}
\end{figure*}
\fi

\section{Anomalous WHOIS Records}
\label{app:anomalous}

There are several classes of anomalies within
the RIR WHOIS records that bear note.  First, there exist duplicate
prefixes within ARIN; this seems to be most common due to
organizational changes, e.g., Akamai acquiring
Linode~\cite{akamai-linode}.
We find no duplicate prefixes for other RIRs.

Second, the use of ``inetnum'' and ``NetRange,'' where the record has
a starting and ending IP address, permits ranges that cannot be
covered within a single contiguous power-of-two prefix. ARIN, RIPE,
APNIC, and \afrinic have
17.8k, 33.1k, 21.2k and 857
of these non-power-of-two WHOIS records respectively.

\begin{table*}[t!]
\caption{Alignment of WHOIS prefixes within the BGP}
\label{tab:bgp}
\begin{tabular}{l rr rr rr rr rr rr}
\toprule
   WHOIS$\in$BGP & \multicolumn{2}{c}{ARIN}
                 & \multicolumn{2}{c}{RIPE}
                 & \multicolumn{2}{c}{APNIC}
                 & \multicolumn{2}{c}{LACNIC}
                 & \multicolumn{2}{c}{AFRINIC}
                 & \multicolumn{2}{c}{ALL} \\
\cmidrule(lr){2-3}
\cmidrule(lr){4-5}
\cmidrule(lr){6-7}
\cmidrule(lr){8-9}
\cmidrule(lr){10-11}
\cmidrule(lr){12-13}
	 & IPv4 & IPv6 & IPv4 & IPv6 & IPv4 & IPv6 & IPv4 & IPv6 & IPv4 & IPv6 & IPv4 & IPv6\\
\midrule
  Subnet & 96.0\% & 95.5\% & 95.6\% & 92.0\% & 92.6\% & 27.7\% & 93.0\% & 50.5\% & 84.8\% & 94.2\% & 95.0\% & 84.5\%\\
 Aligned & 1.9\% & 1.7\% & 2.5\% & 2.7\% & 3.0\%      & 8.7\% & 2.6\% & 18.6\% & 8.9\% & 2.1\% & 2.5\% & 3.5\%\\
Supernet & 0.4\% & 0.4\% & 0.6\% & 0.5\% & 1.7\%      & 1.4\% & 2.9\% & 13.3\% & 2.4\% & 0.5\% & 0.9\% & 1.0\%\\
 MixedAS & 0.2\% & 0.2\% & 0.2\% & 0.3\% & 0.4\%      & 0.4\% & 0.3\% & 0.5\% & 0.3\% & 0.1\% & 0.3\% & 0.3\%\\
Unadvertised & 1.5\% & 2.2\% & 1.1\% & 4.5\% & 2.3\%      &61.8\% &1.2\%& 17.1\% & 3.6\% & 3.1\% & 1.3\% & 10.7\% \\
\bottomrule
\end{tabular}
\end{table*}

Third, in the ``status'' field, which indicates e.g., if a 
prefix is allocated or assigned, we find variation in the
value's textual representation.  For example, in APNIC
there are nine representations for ``Assigned
Non-portable'' with different variations of capitalization and
spacing and even ``typo capitalizations'' (e.g.,
``ASSIGNEd NON-PORTABLE'').  We surmise that such 
oddities are a result of manual record entry---a surprising
finding given the number of records and the widespread use of the WHOIS, requiring accurate records.

We ignore pathologies in the RIR data, the most
common of which are circular references.  For example, an RIR may
list a prefix as transferred to a different RIR, while that RIR
lists the same prefix as belonging to the original RIR.  While these
types of errors could be due to time differences in the data dumps,
the ``Updated'' timestamps on the records suggest they are 
simply errors that we cannot resolve.

Finally, there is inter-RIR inconsistency in record fields and values.
For instance, RIPE records use ``ASSIGNED PA'', APNIC uses
``ASSIGNED PORTABLE'', while LACNIC uses ``assigned.'' Such differences
highlight the organic and ad-hoc way in which RIRs have evolved, and
significantly complicate not only data processing, but a cohesive
understanding of the ecosystem.  Initiatives such as the IETF's RDAP
\cite{RFC7483}, which promise to move the WHOIS to a more
standards-based format both for the data representation and protocol,
will be a welcome change to facilitate future research.

\section{BGP Alignment}
\label{app:bgp}

Table~\ref{tab:bgp} shows the fraction of prefixes for each RIR that 
account for the five BGP-to-WHOIS alignment possibilities
(see \Cref{fig:bgp} for a graphical representation of these five cases).

In IPv6, the four classes are similarly prominent as in IPv4, while we see much more diversity between RIRs.
In APNIC for example, we only have 27.7\% of all prefixes in WHOIS as subnets of BGP, with more than 60\% of all IPv6 APNIC WHOIS entries not being found in BGP at all.
Upon investigation, we find that a single entity---the Indian company Honesty Net Solutions \cite{hns}---is responsible for 79\% of all entries in APNIC's WHOIS database missing in BGP.
These are all /48 subprefixes of \texttt{2401:4800::/32} and have almost all been last updated in WHOIS in May 2012.

\section{Speed-of-Light Sensitivity}
\label{app:sol}

As our method is relies on latency-based constraint based
geolocation, a potential source of error are measurement nodes that
produce latencies larger than expected given their physical location,
\eg due to circuitous routing, last-mile effects, or non-traditional 
media such as satellite links.  Among the RIPE Atlas nodes used across our measurements, we
identified five that reside within Starlink's autonomous system,
ASN14593.  Recall that each prefix is measured by 20
different Atlas nodes.  Among the prefixes measured in our
study, our system selected a Starlink node as one of the 20 active
measurements for 138 prefixes.  We re-ran our analysis to exclude these
Starlink Atlas probes/results (while keeping the other 19
probes per prefix) and the overall macro results remain unchanged,
\ie none of the Starlink nodes were ever selected as the minimum-RTT node
by which to base the Constraint-Based Geolocation (CBG) constraints.  

To examine the sensitivity of our results to the selected speed of
light parameter, we re-ran our CBG analysis to compare 2/3c
versus 1c.  We find that using 1c increases the number of fully
consistent prefixes by approximately 0.1\%.

\label{page:last}

\end{document}